\documentclass[prl,preprint,nofootinbib]{revtex4}

\usepackage{epsf}

\begin{document}

\title
{\bf Pseudogap in High - Temperature Superconductors}
\author{M.V.Sadovskii}

\affiliation
{Institute for Electrophysics, \\
Russian Academy of Sciences, Ural Branch,\\
Ekaterinburg 620016, Russia\\
E-mail:  sadovski@iep.uran.ru}


\begin{center}
{\sl To appear in Physics Uspekhi}
\end{center}

\begin{abstract}

This review describes the main experimental facts and a number of theoretical
models concerning the pseudogap state in high - temperature superconductors.
On the phase diagram of HTSC - cuprates the pseudogap state is observed in the
region of current carrier concentrations less than optimal and leads to a
number of anomalies of electronic properties. These pseudogap anomalies
apparently are due to fluctuations of antiferromagnetic short - range order
developing as system moves towards antiferromagnetic region on the phase
diagram. Electron interaction with these fluctuations leads to anisotropic
renormalization of electronic spectrum and non - Fermi liquid behavior on
certain parts of the Fermi surface. We also discuss some simple theoretical
models describing the basic properties of the pseudogap state, including
the anomalies of superconducting state due to pseudogap renormalization of
electronic spectrum.

\end{abstract}


\maketitle

\newpage

\tableofcontents

\newpage

\section{Introduction.}

The studies of high - temperature superconductivity (HTSC) in copper oxides
remains one of the central topics of the condensed - matter physics.
Despite of the great effort of both experimentalist and theorists the nature of
this phenomenon is still not fully understood. It is common belief that the
basic difficulties here are due to quite unusual properties of these systems in
the normal (non - superconducting) state, while without the explanation of 
these anomalies it is hard to expect the complete understanding of microscopic
mechanism of HTSC. In recent years one of the main directions in physics of HTSC
were the studies of the so called pseudogap state \cite {BatlVar}, which is
observed in the region of the phase diagram, corresponding to current carrier
concentrations less than the optimal concentration (i.e. that corresponding to
the maximal critical temperature of superconducting transition $T_c$). This is
usually called the ``underdoped'' region. In this region a number of anomalies
of electronic properties is observed both in normal and superconducting states,
which are interpreted as due to a drop of the density of single particle states
and anisotropic renormalization of the spectral density of current carriers.
Understanding of the nature and properties of the pseudogap state remains the
central problem in any approach towards explaining the complicated phase 
diagram of HTSC - systems. This problem was already discussed in some 
hundreds of experimental and theoretical papers\footnote{Note that only the
{\em cond-mat} electronic preprint archive contains more than 600 papers somehow
related to the physics of the pseudogap state, while at the last major
$M^2S-HTSC-VI$ Superconductivity Conference in Houston (February 2000) 
four oral sessions were devoted to this problem -- more than to any other
problem in the field of HTSC.}. 

The aim of this review is to report basic experimental facts on the observation
of the pseudogap state in underdoped HTSC -- cuprates, as well as the 
discussion of a number of simple theoretical models of this state.
The review is not thorough, both in describing the experimental data and also
in theoretical part. Experiments are discussed rather briefly, taking into
account the existence several good reviews \cite {Tim,R,RC,TM,TL}.
Theoretical discussion is also rather incomplete and reflects mainly the
personal position of the author. There exist two main theoretical scenarios
for the explanation of pseudogap anomalies in HTSC -- systems. The first is
based upon the model of Cooper pairs formation already above the critical
temperature of superconducting transition \cite {Gesh,EK,Lev,Lok}, 
with phase
coherence appearing for $T<T_c$. 
The second one assumes that the appearance of
the pseudogap state is due to fluctuations of some short -- range order of
``dielectric'' type which exist in the underdoped region of the phase
diagram. Most popular here is the picture of antiferromagnetic (AFM)
fluctuations \cite{Kam,Benn,Dei,Sch,KS}, while the similar role of fluctuations
of charge density waves (CDW), structure or phase separation on microscopic
scale can not be excluded. The author believes that some recent experiments
strongly suggest that this second scenario is correct. Accordingly, in
theoretical part of this review we limit ourselves only to discussion of
this kind of models with detailed enough explanation of the results obtained
by the author and his collaborators, as well as some other authors whose
works are somehow close to ours in main ideas and in approaches used.
Thus, we mainly assume the relevance of the model of antiferromagnetic
fluctuations, though it is still can not be considered as fully confirmed.

The references are also apparently incomplete, moreover these do not reflect
any considerations of priority. It is assumed that the reader will find the
additional references in papers cite here. The author expresses his excuses in
advance to those numerous authors, whose papers are not cited here, mainly not
to make the reference list too long.

\section {Basic Experimental Facts.}

\begin{figure}
\epsfxsize=8cm
\epsfysize=8cm
\epsfbox{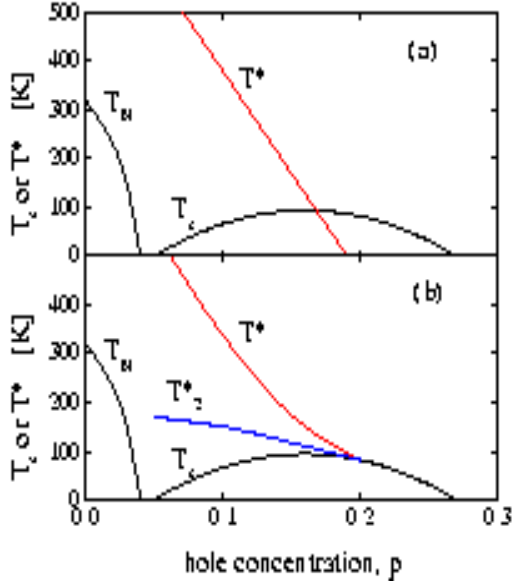}
\caption{Two possible types of the phase diagram of HTSC -- cuprates.}
\label{figpd}
\end{figure}

Typical variants of the phase diagram of HTSC -- cuprates are shown in
Fig.\ref{figpd}.
Depending on concentration of current carriers (in most cases holes) in
highly conducting $CuO_2$ plane a number of phases is observed as well as 
regions with anomalous physical properties. For small concentration of holes
all known HTSC -- cuprates are antiferromagnetic insulators.
As hole concentration grows Neel temperature $T_N$ drops rapidly from the values 
of the order of hundreds of $K$, becoming zero at hole concentration 
$p$ of the order of $0.05$ or even less, while the system becomes metallic 
(though this metal is rather bad). With the further growth of hole concentration
the system becomes superconducting and superconducting critical temperature
at first grows with hole concentration, becoming maximal at 
$p_0\approx 0.15-0.17$ (optimal doping), then drops to zero at
$p\approx 0.25-0.30$, while in this (overdoped) region the metallic behavior
persists. Metallic properties in the region of $p>p_0$ are more or less
traditional (Fermi - liquid), while for $p<p_0$ the system is a kind of 
anomalous metal, not described (in the opinion of majority researchers)
by Fermi - liquid theory\footnote{Note that the question of presence or
absence of Fermi - liquid behavior in HTSC -- systems is rather obscured
in discussions of numerous authors standing on opposite positions and even
using different definitions of Fermi - liquid. In the following we shall use
this terminology according to the opinion of rather badly defined
``majority''. }.

Anomalies of physical properties which are attributed to the appearance of the
pseudogap state are observed in metallic phase for $p<p_0$
 and temperatures
$T<T^*$, where $T^*$ falls from temperatures of the order of $T_N$
for $p\sim 0.05$, becoming zero at some ``critical'' concentration 
$p_c$, which is may be slightly larger than $p_0$ (Fig.\ref{figpd}(a)), e.g.
according to Ref.\cite{TL} this happens at  $p=p_c\approx 0.19$. 
In the opinion of a number of authors (mainly supporters of superconducting
scenario of pseudogap formation) $T^*$ gradually approaches the line of
superconducting $T_c$ close to the optimal concentration $p_0$ 
(Fig.\ref{figpd}(b)). Below we shall see that the majority of new experimental
data apparently confirm the type of the phase diagram shown in
Fig.\ref{figpd}(a) (see details in Ref. \cite{TL}).
 It should be stressed that
temperature $T^*$, in the opinion of majority of researchers, does not have the
meaning of a temperature of some kind of phase transition. It only determines
some characteristic temperature scale below which the pseudogap anomalies 
appear in different measurements. Any kind of singularities of thermodynamic
characteristics in this region of phase diagram are just absent
\footnote{Note, however, an opposite opinion expressed in a recent paper 
\cite{CLMN}, where the line of $T^*$ on the phase diagram was directly
connected with some ``hidden'' 
symmetry breaking, strongly smeared by internal
disorder.}. The general statement is that all these anomalies can be related
to some drop (in this region) in  the density of single -- particle
excitations close to the Fermi level, which corresponds to a general concept of
the pseudogap
 \footnote{The concept of pseudogap was first formulated by Mott
in his qualitative theory of disordered (non -- crystalline) semiconductors
\cite{Mott}. According to Mott the term ``pseudogap'' describes a region of
diminished density of states within the energy interval corresponding to a
band gap in an ideal crystal, representing the ``reminiscence'' of this
band gap which is conserved after strong disordering (amorphization, melting
etc.)}. In this case the value of $T^*$ is just proportional to energy
width of the pseudogap. Sometimes an additional temperature scale
$T^*_2$ is also introduced, as shown in Fig. \ref{figpd}(b), attributed to a
crossover from ``weak'' to ``strong'' pseudogap regime \cite{Sch}, 
usually referring to some change in spin system response in the vicinity of
this temperature. In the following we shall not be dealing with these details.

Now let us describe the most typical experimental signatures of pseudogap
anomalies in HTSC -- cuprates.

\subsection {Specific Heat and Tunneling.}

\begin {figure}
\epsfxsize=9cm
\epsfysize=11cm
\epsfbox{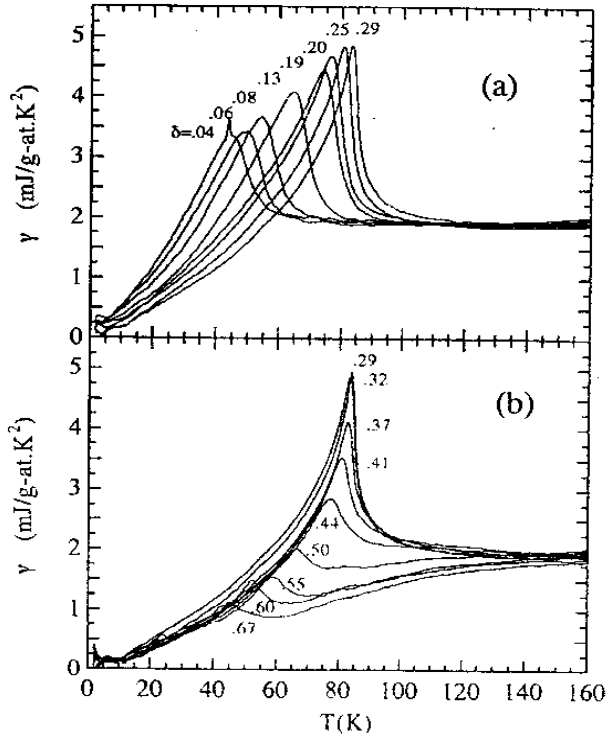}
\caption {Electronic specific heat coefficient $\gamma$ in overdoped
(a) and underdoped (b) $Y_{0.8}Ca_{0.2}Ba_2Cu_3O_{7-\delta}$ 
\cite{Loram}.}
\label{Cybco}
\end {figure}

Consider first the data on electronic specific heat of HTSC -- cuprates.
In metals this contribution is usually represented as $C=\gamma(T)T$, so that
in the normal state ($T>T_c$) $\gamma\sim N(0)$, where $N(0)$ -- is the density
of states at the Fermi level. For $T=T_c$ we have the well known anomaly due
the second order transition and the value of $\gamma(T)$ acquires the
characteristic peak (``jump'' or discontinuity). 
As an example in Fig.\ref{Cybco} 
we show some typical data obtained for 
$Y_{0.8}Ca_{0.2}Ba_2Cu_3O_{7-\delta}$ with different values of $\delta$
\cite{Loram}.
In optimally doped and overdoped samples $\gamma(T)$ remains practically
constant for all $T>T_c$, while for underdoped samples a significant drop of
$\gamma(T)$ is observed for temperatures $T<150-200K$. This observation directly
signifies to the drop in the electronic density of states at the Fermi level
and formation of the pseudogap for  $T<T^*$.

Note also that the value of the ``jump'' of specific heat at $T_c$ is much
reduced as we go into the underdoped region. 
More detailed analysis \cite{TL} 
shows that the appropriate 
value of $\Delta\gamma_c$ drops rather sharply,
starting from the ``critical'' concentration of carriers $p_c\approx 0.19$, 
which is considered to be the point of appearance of the pseudogap in 
electronic spectrum.

\begin {figure}
\epsfxsize=9cm
\epsfysize=8cm
\epsfbox{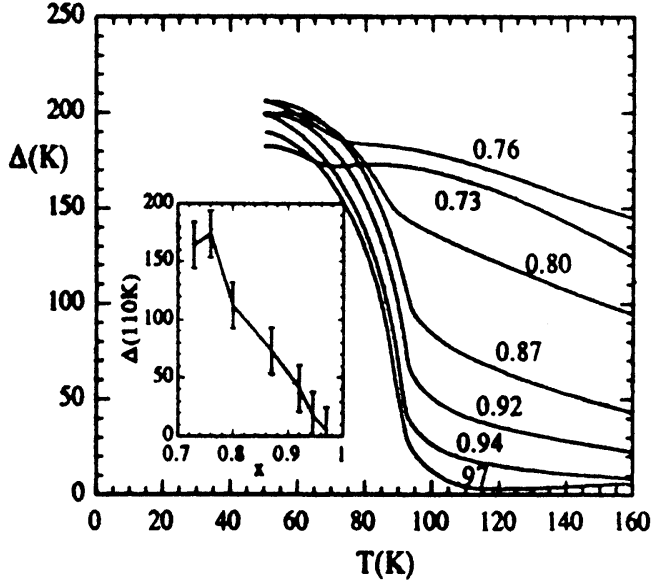}
\caption{Temperature dependence of the energy gap for samples of
$YBa_2Cu_3O_{6+x}$ with different compositions, determined from specific heat
measurements \cite{Lor}.
 The number at different curves denote the 
appropriate values of oxygen content $x$.}
\label{DcT}
\end {figure}

Using more or less traditional expressions of BCS theory, specific heat data
can be used to estimate the value of superconducting energy gap in electronic
spectrum and its temperature dependence.
In Fig.\ref{DcT} we show the data of Ref.\cite{Lor}, obtained for
$YBa_2Cu_3O_{7-\delta}$ ($YBCO$). It is seen that thus defined energy gap
does not go to zero for $T=T_c$ (as it should be in a traditional
superconductor) and there is a significant ``tail'' in the region of higher
temperatures. This effect is stronger for more underdoped samples.
Often these data are naively interpreted as signifying Cooper pair formation
at temperatures $T>T_c$.

Pseudogap formation is clearly seen in the experiments on single particle
tunneling. In a widely cited Ref.\cite{Renn}
 tunneling measurements were
performed on single crystals of $Bi_2Sr_2CaCu_2O_{8+\delta}$ ($Bi-2212$) 
with different oxygen content. For underdoped samples the pseudogap
formation in the density of states was clearly observed for temperatures
higher than $T_c$.
 This pseudogap smoothly evolved into superconducting gap for
$T<T_c$, which again was seen by many as a direct confirmation of
superconducting nature of the pseudogap. Some signs of pseudogap existence was
observed in this work also for slightly overdoped samples.

\begin{figure}
\epsfxsize=9cm
\epsfysize=9cm
\epsfbox{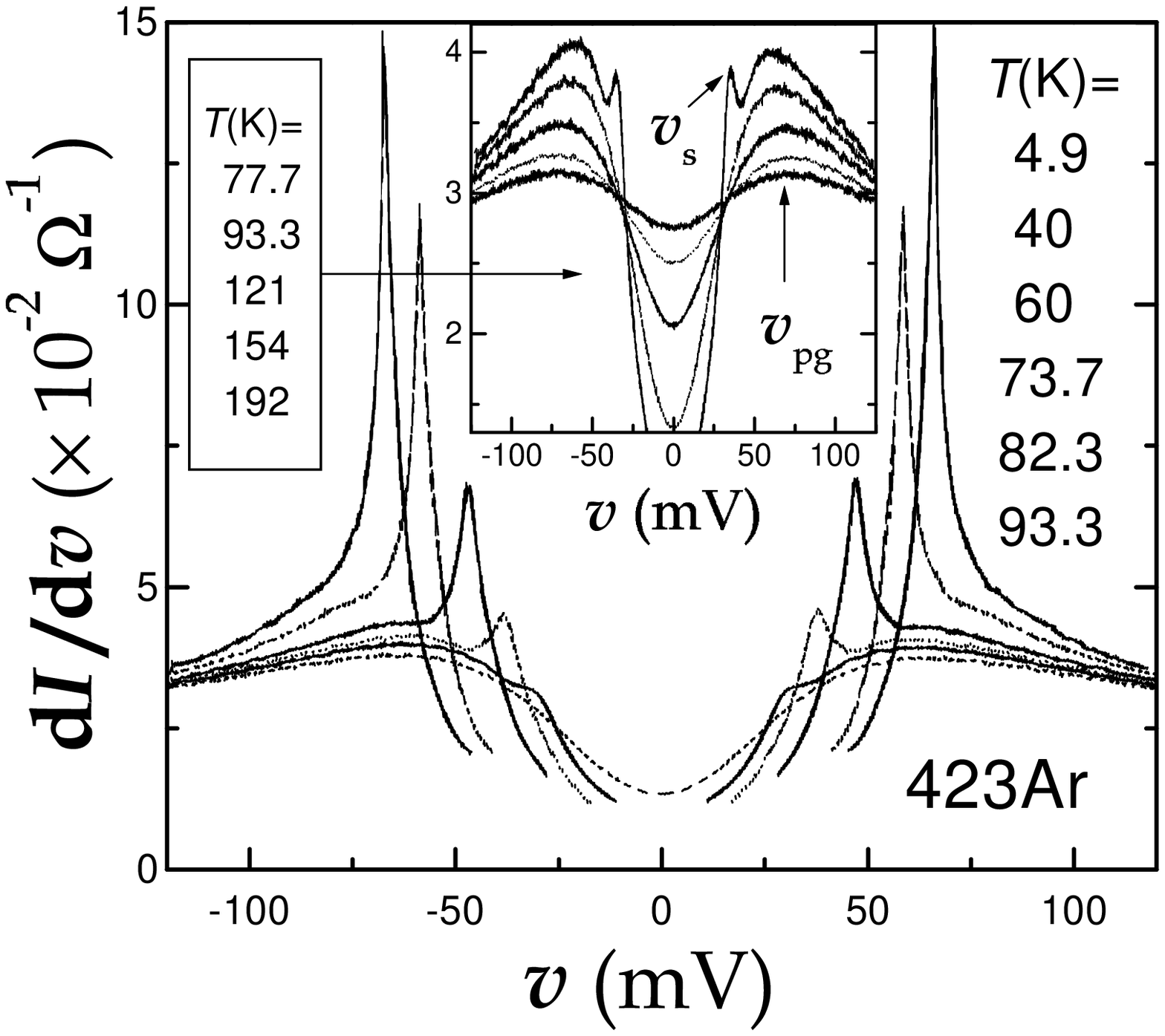}
\caption {Differential conductance of nearly optimally doped 
$Bi-2212$ for different temperatures. At the insert -- results for higher $T$
\cite{Kras1}.}
\label{tunn}
\end{figure}

\begin{figure}
\epsfxsize=9cm
\epsfysize=9cm
\epsfbox{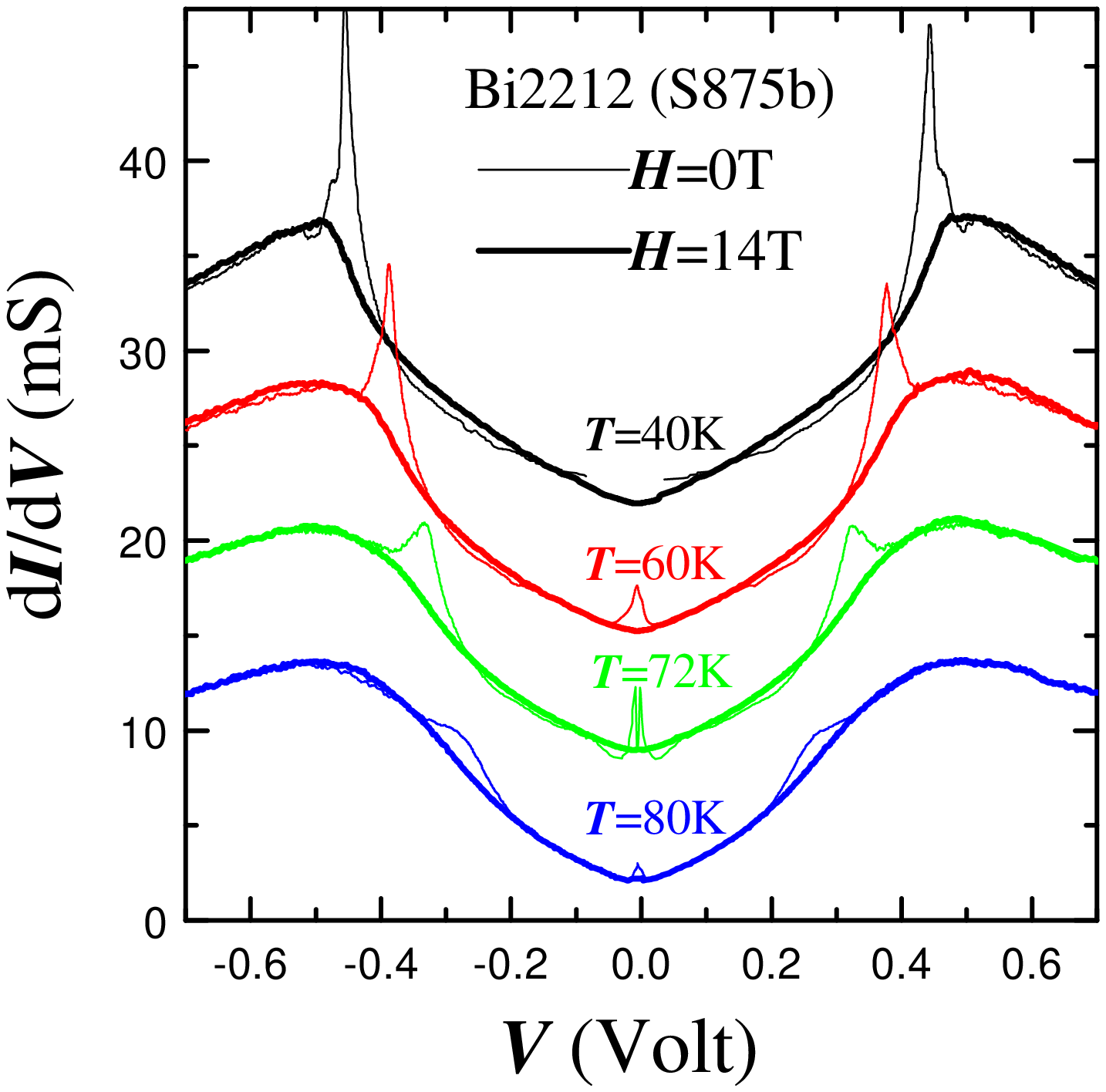}
\caption {Differential conductance of nearly optimally doped $Bi-2212$ 
for different temperatures in magnetic field $H=0$ (thin lines) and for
$H=14T$ (thick lines) \cite{Kras2}.}
\label{tunnH}
\end{figure}

While discussing tunneling experiments in HTSC -- cuprates, there is always 
a question of the surface quality of the samples. Thus, especially interesting
the recent studies of Refs. \cite{Kras1,Kras2}, where an intrinsic tunneling
was measured on the so called mesa -- structures
\footnote{``{\em mesa} (Sp.), a land formation having a flat top and steep
rock walls: common in arid and semiarid parts of the U.S. and Mexico''
(The Random House Dictionary of the English Language. Random House, Inc., 
1969.).}, specially formed (by microlitography) on the surface of the same 
system $Bi-2212$. These experiments clearly demonstrated the existence of
superconducting gap approaching zero at 
$T=T_c$ on the background of smooth
pseudogap, which exists at higher temperatures.
The appropriate tunneling date are shown in Fig.\ref{tunn}.
Typical peaks due to superconducting gap opening are seen on the background
of smooth minimum in the density of states due to pseudogap.
It is especially important that in Ref.\cite{Kras2} it was demonstrated that
superconducting peculiarities of tunneling characteristics were destroyed
by the external magnetic field, while the pseudogap remained practically
independent of the field strength, which clearly shows its non --
superconducting nature. Appropriate data are shown in Fig.\ref{tunnH}

\subsection {NMR and Transport Properties.}

Formation of the pseudogap signals itself also in transport properties of
HTSC -- systems in normal state, Knight shift and NMR -- relaxation.
In particular it is thought to be a reason for the change of the standard 
(for optimally doped samples) linear temperature dependence of resistivity
in the region of $T<T^*$ for underdoped samples. The value of the Knight shift 
in underdoped samples becomes temperature dependent -- for $T<T^*$ it drops 
rather rapidly. Analogous behavior is observed in underdoped region for the 
value of $(TT_1)^{-1}$, where $T_1$ -- is NMR relaxation time. Let us remind
that in usual metals the Knight shift is proportional to  $N(0)$, while
$(TT_1)^{-1}\sim N^2(0)$ (Korringa relation). Resistivity $\rho$ is usually
proportional to the scattering rate (inverse mean -- free time)
$\gamma\sim N(0)$. Thus, the notable suppression of these characteristics
is naturally explained by the drop in the density of states $N(0)$ at the Fermi
level\footnote{Historically the first data on the significant drop of the
density of states in underdoped cuprates were obtained in both NMR and 
magnetic neutron scattering. Accordingly at this initial stage the term 
``spin gap'' was introduced and actively used. However, from further studies it
became clear that the analogous effects are observed not only for spin degrees
of freedom and the term ``pseudogap'' became widely accepted.}. 
Note that this interpretation is obviously oversimplified, moreover when we
discuss the temperature dependence of these physical characteristics.
For example, in the case of resistivity this dependence is usually mainly 
determined by inelastic scattering processes and the physical mechanism of 
these in HTSC -- cuprates is still far from being clear. The drop in the 
density of states, interpreted as partial dielectrization of electronic
spectrum, may also lead to the appropriate growth of resistivity.

\begin{figure}
\epsfbox{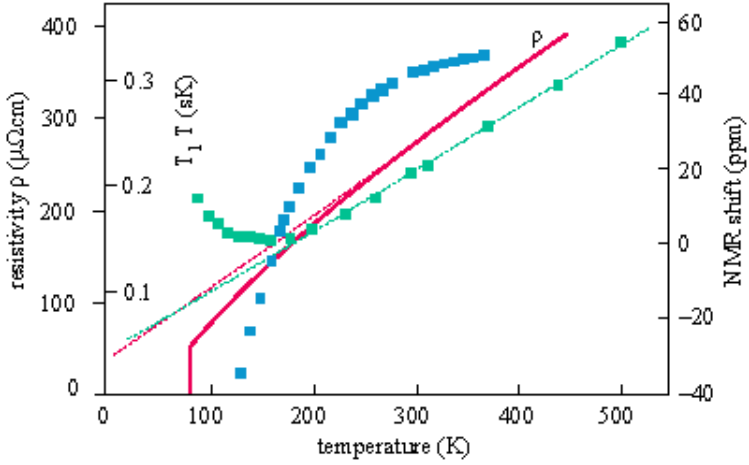}
\epsfxsize=10cm
\epsfysize=9cm
\caption{Influence of the pseudogap formation on resistivity, Knight shift
and NMR relaxation in underdoped $YBa_2Cu_4O_8$. The changes in temperature
dependences in the region of $T<200-300K$ is attributed to pseudogap
formation in the density of states \cite{BatlVar}.}
\label{KTR}
\end{figure}

\begin{figure}
\epsfxsize=9cm
\epsfysize=6cm
\epsfbox{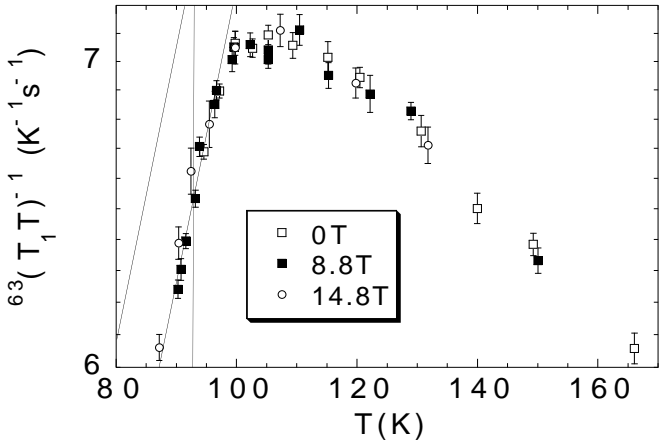}
\caption {NMR relaxation rate $(TT_1)^{-1}$ for slightly underdoped
$YBa_2Cu_3O_{7-\delta}$ in strong magnetic field, directed along the
$c$-axis of the sample \cite{Gorn}. Vertical line shows $T_c$ in zero field
.
The straight line parallel to the data estimates the expected shift of data
in the case of pseudogap suppression by magnetic field, its position is
determined by the observed shift of $T_c$.}
\label{T1H}
\end{figure}

In Fig.\ref{KTR} taken from Ref.\cite{BatlVar} we show the summary of 
experimental data for underdoped samples of $YBa_2Cu_4O_8$ using the
results of Refs.\cite{Buch,Yas,All}.
In Fig.\ref{T1H} we show the data of Ref.\cite{Gorn} on the measurements of
$(TT_1)^{-1}$ on $^{63}Cu$-nuclei in slightly underdoped 
$YBa_2Cu_3O_{7-\delta}$ in strong enough magnetic field.
It is seen that the magnetic field dependence of the data is practically
absent, giving the strong support for non - superconducting nature of the
pseudogap, as in the opposite case magnetic field should significantly change
the size of the effect. Note that the expected influence of magnetic field 
upon the Knight shift and NMR relaxation of $^{63}Cu$ was observed in slightly
overdoped $TlSr_2CaCu_2O_{6.8}$ where it was accurately described by the effects
of suppression of superconducting fluctuations \cite{Guo}. This fact can be seen
as an evidence of disappearance of the pseudogap of non - superconducting
nature (independent of magnetic field) at some lower concentration of current
carriers.

Let us stress that in different experiments the characteristic temperature
$T^*$ below which the anomalies attributed to pseudogap appear can somehow 
change
 depending on the quantity being measured. However, in most cases there
is a systematic dependence of $T^*$ on doping and this temperature tends to
zero close to optimal concentration of current carriers or at slightly higher
concentrations.
\begin{figure}
\epsfxsize=8cm
\epsfysize=8cm
\epsfbox{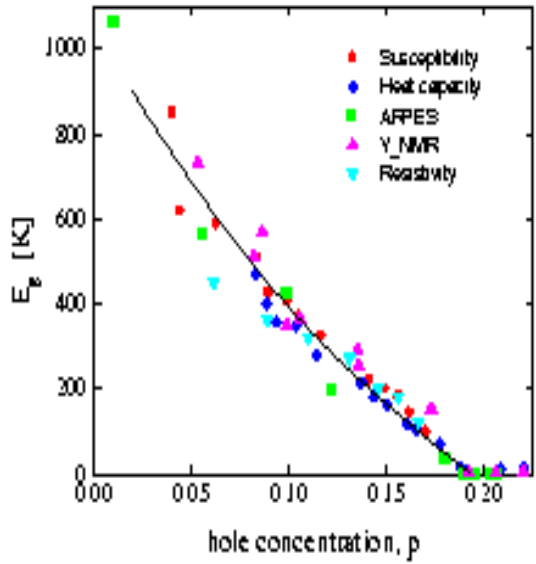}
\caption {Dependence of energy scale characterizing the width of the pseudogap
$E_g$ in $YBCO$
 on concentration of holes following from different
experiments \cite{TL}.}
\label{Eg}
\end{figure}
It is clear that $T^*$ does not represent any strictly defined temperature
(e.g. of some phase transition) but only determines some characteristic
crossover temperature (energy) scale below which pseudogap anomalies appear .
This energy scale may be determined from experimental data using some rough
model of the energy dependence of the density of states close to the Fermi level
within the pseudogap region (e.g. triangular with the width of $E_g$,
giving $T^*=0.4E_g$) \cite{TL}. In Fig.\ref{Eg} we present the summary of the
values of pseudogap width determined in this way from different experiments on
$YBCO$ for different hole concentrations \cite{TL}. It is seen that the
pseudogap ``closes'' at some critical value of $p_c\approx 0.19$. Once again
let us stress that both the line of $T^*$ on the phase diagram and ``critical''
concentration $p_c$ apparently possess only qualitative meaning. However,
there are also attempts to interpret $p_c$ as some kind of a ``quantum''
critical point \cite{TaLor}.

\subsection {Optical Conductivity.}

\begin{figure}
\epsfxsize=9cm
\epsfysize=8cm
\epsfbox{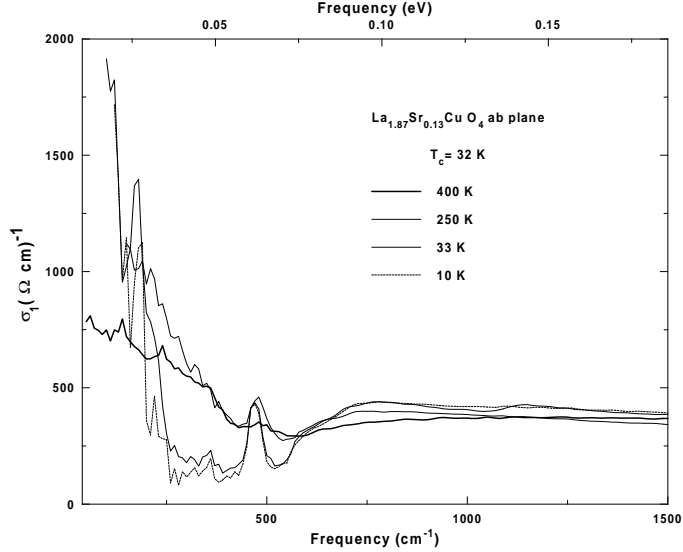}
\caption{Real part of optical conductivity in $ab$ - plane of  
$La_{1.87}Sr_{0.13}CuO_4$ for different temperatures \cite{STP}.}
\label{optcond}
\end{figure}

Pseudogap formation in underdoped HTSC -- cuprates is also clearly seen in
numerous measurements of optical conductivity, both for electric field vector
polarized along the highly conducting $CuO_2$ plane and along orthogonal
direction of tetragonal axis $c$. These experiments are reviewed in detail
in Ref.\cite{Tim,PBT}. As a typical example in Fig.\ref{optcond} we show the 
data of Ref.\cite{STP} on optical conductivity in the $CuO_2$
 - plane of
underdoped $La_{1.87}Sr_{0.13}CuO_4$. 
Characteristic feature of these data is the presence of rather narrow
``Drude - like'' peak in the frequency region $\omega<250 cm^{-1}$ and the
appearance of ``pseudogap suppression'' of conductivity in the interval of
$250-700 cm^{-1}$ accompanied by smooth maximum around $\omega\sim 800 cm^{-1}$. 
Especially sharply pseudogap anomalies are seen after the experimental data
are expressed in terms of the so called generalized Drude formula
\cite{Tim,PBT}, which introduces the effective (frequency dependent)
scattering rate and effective mass of current carriers:
\begin{eqnarray}
\frac{1}{\tau(\omega)}=\frac{\omega_p^2}{4\pi}Re\left(\frac{1}{\sigma(\omega)}
\right) \label{sgdr}\\
\frac{m^*(\omega)}{m}=\frac{1}{\omega}\frac{\omega^2_p}{4\pi}Im\left(
\frac{1}{\sigma(\omega)}\right)
\label{mgdr}
\end{eqnarray}
where $\sigma(\omega)$ -- the observed complex conductivity, $\omega_p$ --
plasma frequency, $m$ -- mass of the free electron.
Note that the use of expressions (\ref{sgdr}),
 (\ref{mgdr}) is just another
form of expression of the same experimental data in the form of 
$1/\tau(\omega)$ and $m^*(\omega)$ instead of two more usual characteristics
--- the real and imaginary parts of conductivity. Accordingly there is no 
especially
 deep physical meaning in both $\tau(\omega)$ and $m^*(\omega)$. 
However, this form of presentation of experimental data in optics is widely
used in the literature.

\begin{figure}
\epsfxsize=9cm
\epsfysize=11cm
\epsfbox{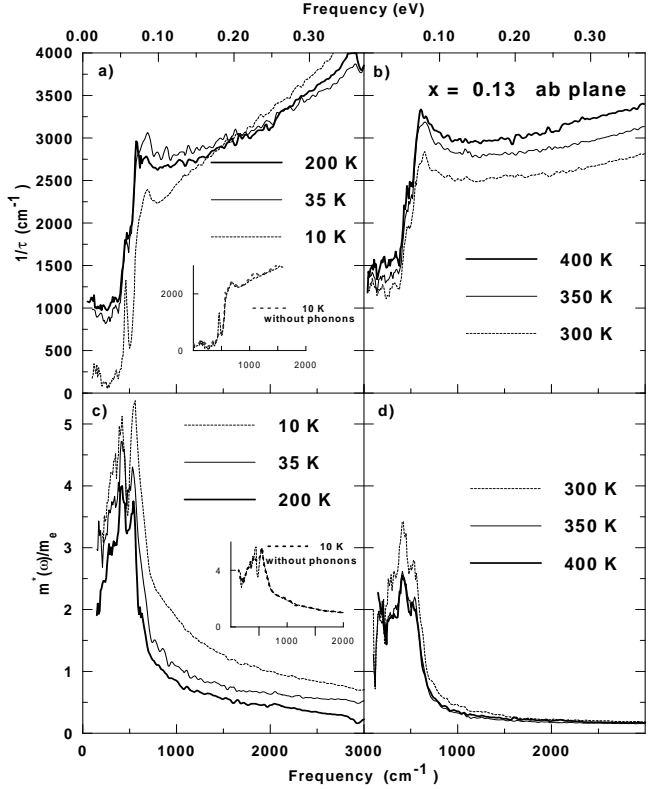}
\caption{Generalized scattering rate and effective mass as functions of the
frequency of the external field obtained from the fit of optical data for
$La_{1.87}Sr_{0.13}CuO_4$ to generalized Drude formula \cite{STP}:
(a), (c) -- for temperatures $T<250K$, when the linear growth of
$1/\tau$ with frequency is observed for $\omega>700cm^{-1}$, 
(b), (d) -- higher temperatures, when the linear growth of
$1/\tau$ is practically invisible.}
\label{genDr}
\end{figure}

In Fig.\ref{genDr} we present the results of such a representation of
experimental data of the sort shown in Fig.\ref{optcond} \cite{STP}. 
It is seen that the effective scattering rate for temperatures less than
$T^*\sim 450K$ 
is sharply suppressed for frequencies of an external field
less than $700 cm^{-1}$, 
while for higher frequencies
$\frac{1}{\tau(\omega)}$ grows linearly with $\omega$,
 demonstrating the
anomalous non - Fermi liquid behavior. The presence of this dip in
$1/\tau(\omega)$ is most vivid reflection of the pseudogap in optical data.
At the same time we have to note that the frequency dependence of
$m^*(\omega)$ shown in Fig.\ref{genDr} apparently has no clear interpretation.

Analogous behavior of $1/\tau(\omega)$ is also observed in data on
transverse optical conductivity $\sigma_c(\omega)$, practically for all
HTSC -- cuprates studied in underdoped region \cite{Tim,PBT}
.

\subsection {Fermi Surface and ARPES.}

Especially spectacular effects due to formation of the pseudogap are
observed in experiments on angle -- resolved photoemission (ARPES) \cite{RC,TM}.
This method is, first of all, decisive in studies of the topology of the Fermi
surface of HTSC -- cuprates \cite{Dessau,Shen,RC}, being practically the sole
source of this information.

\begin{figure}
\epsfxsize=9cm
\epsfysize=9cm
\epsfbox{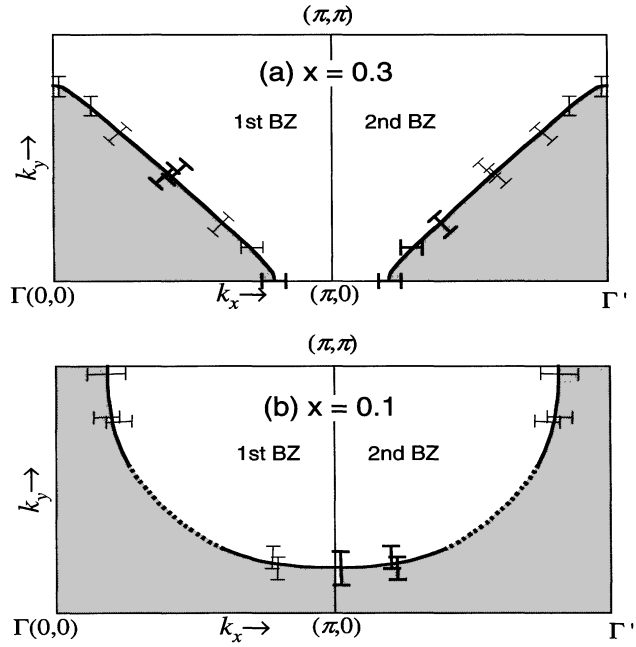}
\caption{Fermi surface of $La_{2-x}Sr_xCuO_4$ for overdoped (a) and
underdoped (b) cases \cite{Ino}.}
\label{FSLSCO}
\end{figure}

\begin{figure}
\epsfxsize=9cm
\epsfysize=9cm
\epsfbox{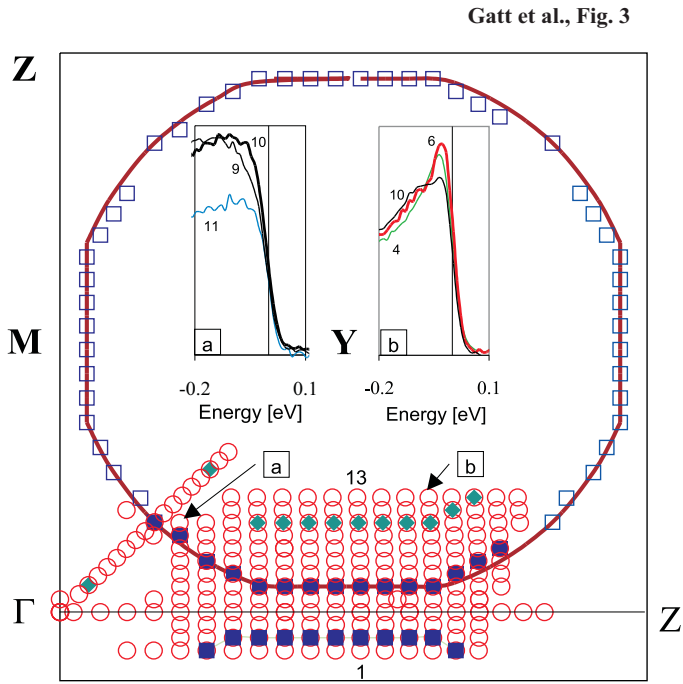}
\caption {Fermi surface of overdoped $Bi-2212$ \cite{Onell}.
 Circles represent
points where measurements were performed. At the inserts -- energy distribution
of photoelectrons along the directions $a$ and $b$, shown by arrows.}
\label{FSBISCO}
\end{figure}

In Fig.\ref{FSLSCO} we show the data of Ref.\cite{Ino} on the Fermi surface of
$La_{2-x}Sr_xCuO_4$ for two different compositions. In overdoped state 
$(x=0.3)$ Fermi surface is electronic and centered around the point
$\Gamma (0,0)$ in Brillouin zone. As concentration $x$ diminishes the topology
of the Fermi surface changes, so that for optimal concentration of $Sr$ and in
underdoped region $(x=0.1)$ it becomes hole - like and centered around the
point $Y (\pi,\pi)$. Precisely this topology is observed in majority of
ARPES -- experiments for almost all HTSC -- cuprates.
As an interesting example in Fig.\ref{FSBISCO} from Ref.\cite{Onell}
 we show
the hole - like Fermi surface of most extensively studied (by ARPES) system
$Bi-2212$ in overdoped state. The spectacular feature here is the observation of
large flat parts of the Fermi surface, orthogonal to direction $YM$ and 
symmetric directions. Independently this result was confirmed in Ref.\cite{ZX} 
for optimally doped system. The presence of the flat parts (patches) on the
Fermi surface may be very important for the construction of microscopic
theories of electronic properties of HTSC -- systems.
Dropping these fine details in the first rough enough approximation the
observed topology of the Fermi surface and the spectrum of elementary
excitations in the $CuO_2$ plane can be rather accurately described by the
tight - binding model:
\begin{equation}
\varepsilon_{\bf k}=-2t(\cos k_xa+\cos k_ya)-4t'\cos k_xa\cos k_ya
\label{tbspectr}
\end{equation}
where $t\approx0.25eV$ -- transfer integral between nearest neighbors, 
and $t'$ -- transfer integral between second nearest neighbors, which may
change between 
$t'\approx -0.45t$ for $YBa_2Cu_3O_{7-\delta}$ and 
$t'\approx -0.25t$ for $La_{2-x}Sr_xCuO_4$, $a$ -- lattice constant.

Quite recently the simple picture described above was seriously revised in
Refs.\cite{ZX,Chu,Gr,Bog} using the new ARPES data obtained with
synchrotron radiation of higher energy, than in previous experiments.
It was claimed that the Fermi surface of $Bi-2212$
 is electronic and
centered around point $\Gamma$. The main discrepancies with previous data
are in the most interesting (for us) vicinity of the point $(0,\pi)$ in the
inverse space. These claims, however, were met with sharp objections from
other groups of ARPES experimentalists \cite{Fr,Mes,Ban,Bor,Leg}. The question
remains open, but in the following we shall adhere to traditional 
interpretation.

There are several good reviews on the observation of pseudogap anomalies in
ARPES  experiments \cite{R,RC,TM}, so below we shall limit ourselves only to
basic qualitative statements. The intensity of ARPES (energy distribution of
photoelectrons) in fact is determined as \cite{RC}:
\begin{equation}
I({\bf k}\omega)=I_0({\bf k})f(\omega)A({\bf k}\omega)
\label{IARP}
\end{equation}
where $\bf k$ -- is the electron momentum in the Brillouin zone,
$\omega$ -- the energy of initial state with respect to the Fermi level
(chemical potential)\footnote{In real experiments $\omega$ is measured with
respect to the Fermi level of some good metal like $Pt$ or $Ag$ which is put
into electric contact with the sample.}, $I_0({\bf k})$ includes kinematic 
factors and the square of the matrix element of electron -- photon
interaction (and in rough enough approximation can be considered a constant).
The value
\begin{equation}
A({\bf k},\omega)=-\frac{1}{\pi}ImG({\bf k},\omega+i\delta)
\label{sdens}
\end{equation}
where $G({\bf k},\omega)$ -- is the Green's function, represents the spectral
density of electrons. Fermi distribution $f(\omega)=[exp(\omega/T)
+1]^{-1}$ 
reflects the fact that only electrons from the occupied states participate in
photoemission. Thus in a rough approximation we may say that ARPES experiments
directly measure the product of $f(\omega)A({\bf k}\omega)$, and we can get the
direct information on the spectral properties of single - particle excitations
in our system.

\begin{figure}
\epsfxsize=9cm
\epsfysize=12cm
\epsfbox{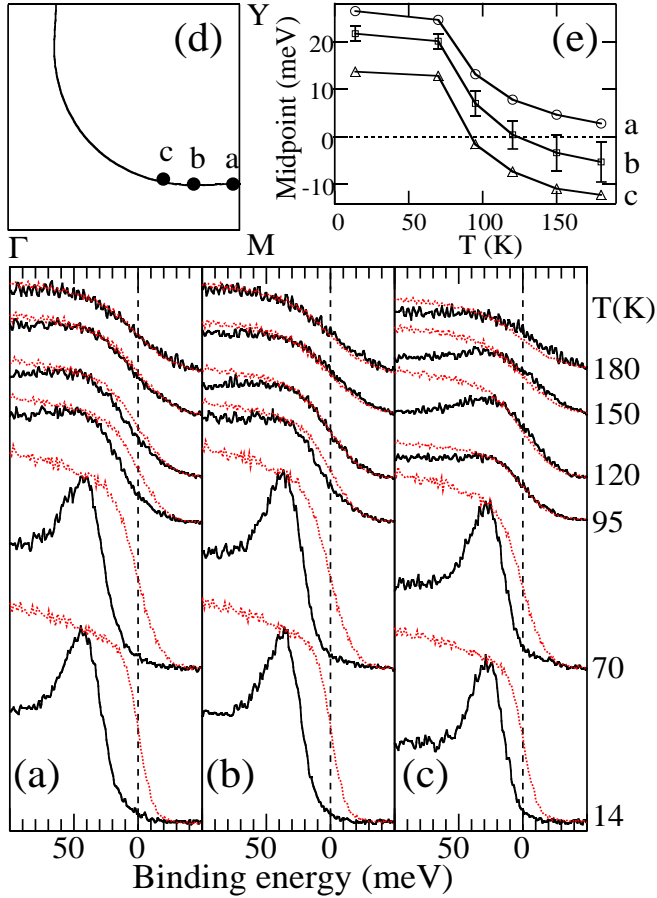}
\caption{ARPES spectra in three different points $a,b,c$ on the Fermi surface of
$Bi-2212$ for underdoped sample with $T_c=85K$. Thin lines -- ARPES spectrum 
of a reference sample of $Pt$ \cite{Norm}.}
\label{ARPES}
\end{figure}

In Fig.\ref{ARPES} we show typical ARPES data for  
$Bi_2Sr_2CaCu_2O_8$ \cite{Norm}, measured in three different points on the 
Fermi surface for different temperatures.
The appearance of the gap (pseudogap) is reflected in the shift (to the left)
of the leading edge of photoelectron energy distribution curve, compared to
the reference spectrum of a good metal $(Pt)$. It is seen that the gap is
closed at different temperatures for for different values of ${\bf k}$, while
its width diminishes as we move away from $(0,0)-(0,\pi)$ direction. 
Pseudogap is completely absent in the direction of the zone diagonal
$(0,0)-(\pi,\pi)$. At low temperatures these data are consistent with the
picture of  $d$-wave pairing, which is well established from numerous
experiments on HTSC -- cuprates \cite{Legg,Iz}.
 It is important, however,
that this ``gap'' in ARPES data is observed also for temperatures significantly
higher than the temperature of superconducting transition $T_c$.

\begin{figure}
\epsfxsize=9cm
\epsfysize=12cm
\epsfbox{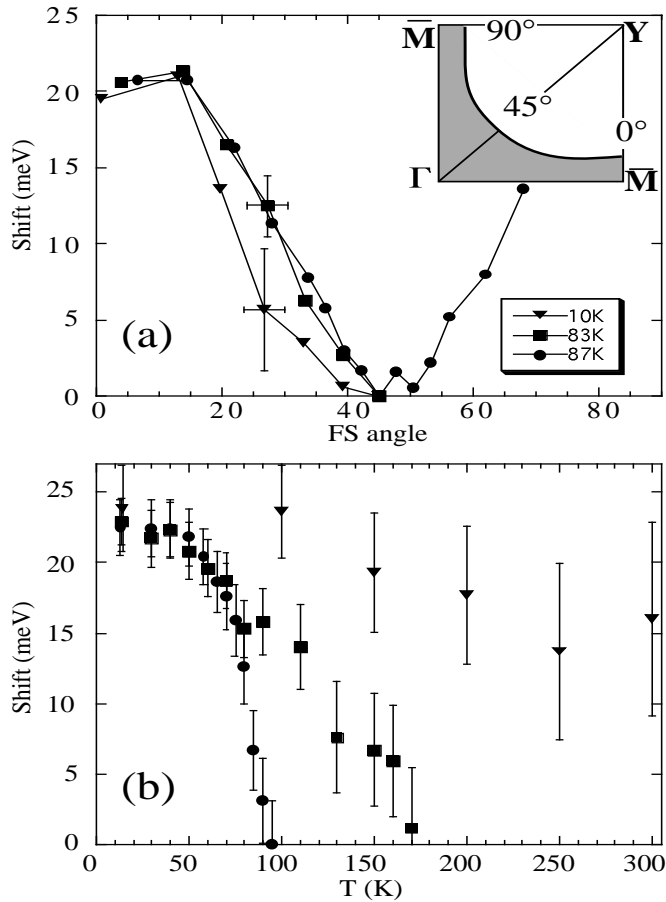}
\caption{Angular (a) and temperature (b) dependence of energy gap in $Bi-2212$,
obtained from ARPES data \cite{Nrm} for samples with  $T_c=87K$ (nearly
optimally doped), $T_c=83K$ and $T_c=10K$ (underdoped). 
}
\label{ARPGAP}
\end{figure}

In Fig.\ref{ARPGAP} we show the angular dependence of the gap width in the
Brillouin zone and the temperature dependence of its maximum value for the
number of samples of $Bi-2212$ with different oxygen content, obtained via
ARPES in Ref.\cite{Din}. It is seen that with the general $d$-wave symmetry,
the gap in the spectrum of optimally doped system becomes zero practically at
$T=T_c$, while for underdoped samples there appear characteristic ``tails'' in
the temperature dependence of the gap for 
$T>T_c$, quite similar to those
shown in Fig.\ref{DcT}.
 Qualitatively we may say that the formation of
anisotropic (in reciprocal space) pseudogap for $T>T_c$, continuously
evolving into superconducting gap for $T<T_c$, leads to a ``destruction'' of the
parts of the Fermi surface of underdoped samples already for $T<T^*$ and around
the point $(0,\pi)$ (and symmetric to it), and that the effective width of these
parts grows as the temperature is lowered \cite{Nrm}. 

\begin{figure}
\epsfxsize=9cm
\epsfysize=9cm
\epsfbox{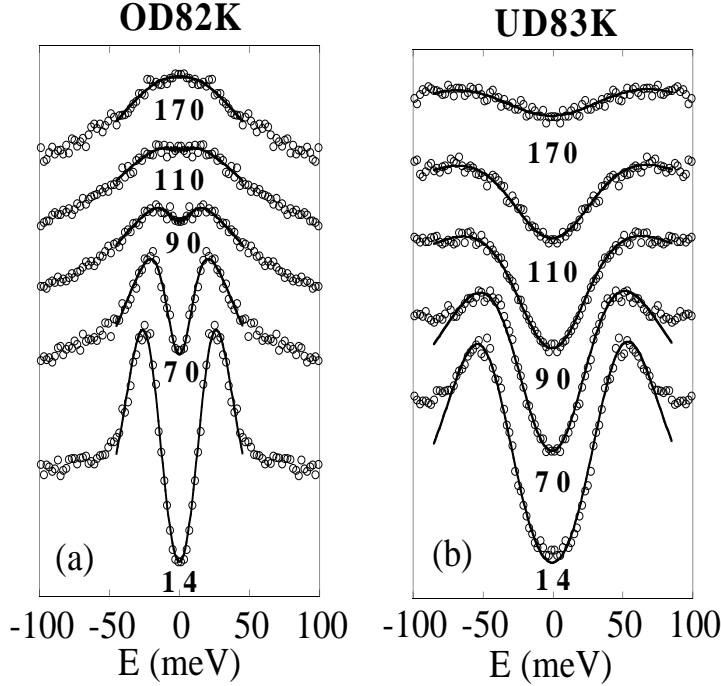}
\caption{Symmetrized ARPES spectra for overdoped sample of $Bi-2212$
 with
$T_c=82K$ (a) and underdoped sample with $T_c=83K$ (b) at the point of
intersection of the Fermi surface with Brillouin zone border
$(0,\pi)-(\pi,\pi)$ \cite{NorRan}. }
\label{Isym}
\end{figure}

Central point is, of course, the evolution of spectral density
$A({\bf k}_F,\omega)$ at the Fermi surface. Under rather plausible assumptions
this may be directly obtained from ARPES data \cite{Nrm}. In case of
electron - hole symmetry we have 
$A({\bf k}_F,\omega)=A({\bf k}_F,-\omega)$
(which is always valid close enough to the Fermi level, in real systems for
$|\omega|$ less than few tenths of $meV$), so that, with the account of 
$f(-\omega)=1-f(-\omega)$, from (\ref{IARP}) for ${\bf k}={\bf k}_F$ we
immediately get  $I(\omega)+I(-\omega)=
A({\bf k}_F,\omega)$. Thus, the
spectral density at the Fermi surface may be obtained directly from
experiments, more precisely from the symmetrized ARPES - spectrum
$I(\omega)+I(-\omega)$. As an example of such approach, in Fig.\ref{Isym}
 we
present the data of Ref.\cite{NorRan} for underdoped sample of $Bi-2212$ with
$T_c=83K$ and overdoped sample with $T_c=82K$ for different temperatures. 
It is seen that the existence of the pseudogap is reflected in typical
two - peak structure of the spectral density, appearing for the underdoped
system at temperatures significantly higher than 
$T_c$.

Let us stress that well defined quasiparticles correspond to sharp enough
peak of the spectral density $A({\bf k}_F,\omega)$ at $\omega=0$. This type
of behavior is practically unobservable in HTSC -- cuprates, at least for
$T>T_c$. In principle, this does not seem surprising -- it is rather
difficult to imagine well defined quasiparticles almost in any system for
$T>100K$! However, the much improved resolution of modern ARPES installations
apparently allows to conclude, that the actual width of the peak observed is
larger than experimental resolution, so that the problem can be studied
experimentally \cite{Kami}. It appears that in superconducting state, for 
$T\ll T_c$, in the vicinity of the crossing of the Fermi surface with the
diagonal of the Brillouin zone ($(0,0)-(\pi,\pi)$ direction), there appears
sharp enough peak of the spectral density, corresponding to well defined
quasiparticles \cite{Kami}. However in the vicinity of the point $(0,\pi)$ 
the Fermi surface is ``destroyed'' by superconducting gap, corresponding
to $d$-wave pairing, which leads to the two - peak structure of the 
spectral density.

The smooth evolution of ARPES pseudogap observed for  $T>T_c$ into
superconducting gap corresponding to $d$-wave pairing for $T<T_c$ is
often considered as an evidence of superconducting nature of the pseudogap.
However, most probably this is wrong. In this respect we should like to
note the very interesting paper \cite{Ronn}, where the ARPES measurements
were performed on dielectric oxide $Ca_2CuO_2Cl_2$, which is structurally
analogous to $La_2CuO_4$ and becomes a high -- temperature superconductor
after doping by $Na$ or $K$. ARPES measurements on this system were possible 
in insulating phase due to especially good quality of the sample surface.
The exciting discovery made in Ref.\cite{Ronn} was apparently first
observation of the ``remnant'' Fermi surface in this Mott dielectric, which
is everywhere ``closed'' by the gap (apparently of the Mott -- Hubbard
nature). At the same time a strong anisotropy of this gap in
inverse space with the symmetry of $d$-wave type was also observed \cite{Ronn},
very similar to analogous data for HTSC -- systems in metallic phase.
It seems plausible that this anisotropic gap has the same nature as 
high - energy pseudogap say in $Bi-2212$ \cite{Ronn}. Of course, it is quite
clear that there is no any kind of Cooper pairing in this insulator.

\subsection {Other Experiments.}

Pseudogap behavior is also observed in a number of other experiments, such as
electronic Raman scattering \cite{Hack} and magnetic scattering of neutrons
\cite{Bourg}. 
All these experiments are interpreted as an evidence for
significant suppression of the single - particle density of states close to the
Fermi level of underdoped HTSC - cuprates for temperatures $T<T^*$, that is
already in normal phase. Due to a space shortage we shall not discuss these 
data in any detail, moreover that a good review of these can be found in papers
cited above, as well as in Ref.\cite{Tim}.

\section {Theoretical Models of Pseudogap State.}

\subsection {Scattering by Fluctuations of Short - Range Order:
Qualitative Picture.}

As was already noted above there exist two main theoretical scenario to explain
pseudogap anomalies of HTSC -- systems.
 The first is based on the model of 
Cooper pair formation above the temperature of superconducting transition
(precursor pairing), while the second assumes that formation of the pseudogap 
state is due to fluctuations of short - range order of ``dielectric'' type
(e.g. antiferromagnetic or charge density wave) existing in the underdoped
region of the phase diagram and leading to strong scattering of electrons 
and
pseudogap renormalization of the spectrum. This second scenario seems more
preferable both because of experimental evidence
\footnote {For example, tunneling experiments of Refs.\cite{Kras1,Kras2}, 
in our opinion, practically rule out ``superconducting'' scenario of pseudogap
formation.
}, described above, and due to the simple fact that all pseudogap anomalies
become stronger for stronger underdoping, when system moves from optimal
(for superconductivity) composition towards insulating (antiferromagnetic) 
region on the phase diagram.

\begin{figure}
\epsfxsize=6cm
\epsfysize=6cm
\epsfbox{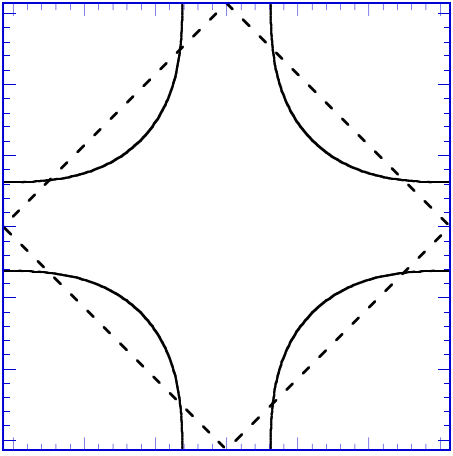}
\caption{Fermi surface in the Brillouin zone and ``hot spots'' model.
Dashed lines represent borders of magnetic Brillouin zone appearing after
period doubling due to appearance of antiferromagnetic long - range order.
``Hot spots'' are defined as points of crossing of the Fermi surface with
borders of magnetic zone.}
\label{hspots}
\end{figure}

Consider typical Fermi surface of current carriers in $CuO_2$ plane, shown
in Fig.\ref{hspots}. Phase transition to antiferromagnetic state leads to
doubling of lattice period and appearance of ``magnetic'' Brillouin zone in
reciprocal space, shown by dashed lines in Fig.\ref{hspots}
.
If the energy spectrum of current carriers is given by (\ref{tbspectr}) 
with $t'=0$, then for the case of half - filled band the Fermi surface is just
a square coinciding borders of magnetic zone and there is complete nesting
 ---
flat parts of the Fermi surface coincide after translation by vector of
antiferromagnetism ${\bf Q}=(\pm\pi/a,\pm\pi/a)$. For $T=0$ electronic spectrum
is unstable --- energy gap is opened everywhere at the Fermi surface and the
system becomes insulator due to formation of antiferromagnetic spin density
wave (SDW)\footnote{The same picture is valid for dielectrization due to
formation of appropriate (period doubling) charge density wave (CDW).}
These arguments lay behind the popular explanation of antiferromagnetism of
HTSC -- cuprates, let us quote e.g. Ref.\cite{SWZ}. The review of this and
similar models can be found in \cite{Izy}. 
In the case of the Fermi surface shown in Fig.\ref{hspots} the appearance of
antiferromagnetic long - range order, in accordance with general rules of the
band theory \cite{Zim}, leads to the appearance of discontinuities of 
isoenergetic surfaces (e.g. Fermi surface) at crossing points with borders of 
new (magnetic) Brillouin zone due to gap opening at points connected by 
vector ${\bf Q}$. 

In the most part of underdoped region of the phase diagram of HTSC -- cuprates
there is no antiferromagnetic long - range order, however quite a number of 
experiments indicates the existence (below the $T^*$ line) of well developed
fluctuations of antiferromagnetic short - range order.
In the model of ``nearly antiferromagnetic'' Fermi liquid \cite{Mont1,Mont2} 
an effective interaction of electrons with spin fluctuations is described
by dynamic spin susceptibility $\chi_{\bf q}(\omega)$, which can be written in
the following form determined from the fit to NMR experiments \cite{MilMon,Iz}:
\begin{equation}
V_{eff}({\bf q},\omega)=g^2\chi_{\bf q}(\omega)\approx
\frac{g^2\xi^2}{1+\xi^2({\bf q-Q})^2-i\frac{\omega}{\omega_{sf}}}
\label{V}
\end{equation}
where $g$ -- is the coupling constant, $\xi$ -- correlation length of spin
fluctuations (short - range order), ${\bf Q}=(\pm\pi/a,\pm\pi/a)$ -- vector of
antiferromagnetic ordering in insulating phase, $\omega_{sf}$ -- characteristic
frequency of spin fluctuations.

Due to the fact that dynamic spin susceptibility  $\chi_{\bf q}(\omega)$ 
possesses peaks at wave vectors smeared around $(\pm\pi/a,\pm\pi/a)$, there
appear ``two types'' of quasiparticles -- ``hot quasiparticles'' with momenta
from the vicinity of ``hot spots'' at the Fermi surface (Fig.\ref{hspots})
and with energies satisfying the following inequality
($v_F$ -- velocity at the Fermi surface):
\begin{equation}
|\varepsilon_{\bf k}-\varepsilon_{\bf k+Q}|<v_F/\xi,
\label{hsp}
\end{equation} 
and ``cold'' ones, with momenta close to the parts of the Fermi surface
around Brillouin zone diagonals $|p_x|=|p_y|$ and not satisfying
(\ref{hsp}) \cite{Sch}. 
This terminology is due to the fact that quasiparticles
from the vicinity of ``hot spots'' are strongly scattered with momentum 
transfer of the order of ${\bf Q}$, because of strong interaction with spin
fluctuations (\ref{V}), while for quasiparticles with momenta far from
``hot spots'' this interaction is much weaker. In the following this model will
be called ``hot spot'' model\footnote{Let us stress that the AFM nature of
fluctuations is in fact irrelevant for further analysis and the model used
in the following is, in fact, more general. The only thing important is the
presence of strong scattering of electrons with momentum transfer from some
narrow vicinity of the vector ${\bf Q}$, which scatters electrons from one
side of the Fermi surface to other. This type of scattering may be also due
to charge (CDW) or structural fluctuations.}.

Correlation length of short - range order fluctuations $\xi$, defined in
(\ref{V}), is very important for the following discussion. Note that in real
HTSC -- systems it is usually not very large and changes in the interval of
$2a<\xi<8a$ \cite{BarzP,P97}. 

Characteristic frequency of spin fluctuations $\omega_{sf}$, depending on the
compound and doping, is usually in the interval of $10-100K$ 
\cite{BarzP,P97}, so that in most part of pseudogap region on the phase
diagram we can satisfy an inequality $\pi T \gg \omega_{sf}$,
 which allows us
to neglect spin dynamics and use quasistatic approximation\footnote{
In terms of Ref.\cite{Sch} this corresponds to the region of ``weak''
pseudogap.}:
\begin{equation}
V_{eff}({\bf q})=\tilde W^2\frac{\xi^2}{1+\xi^2({\bf q-Q})^2}
\label{Vef}
\end{equation}
where $\tilde W$ -- is an effective parameter of dimensions of energy, 
which in the model of AFM fluctuations can be written as \cite{Sch}:
\begin{equation}
\tilde W^2=g^2T\sum_{m{\bf q}}\chi_{\bf q}(i\omega_m)=
g^2<{\bf S}_{i}^{2}>/3
\label{dd}
\end{equation}
where ${\bf S}_i$ is the spin on the lattice site (ion of $Cu$ in the $CuO_2$ 
plane of HTSC -- cuprates). 

In the following we shall consider both $\tilde W$ and $\xi$
 as phenomenological
parameters. In particular, the value of $\tilde W$ determines effective width
of the pseudogap. The full microscopic theory of the pseudogap state will not
be our aim, we shall only try to model the appropriate renormalization of
electronic spectrum and its influence on physical properties, e.g. 
superconductivity.

Significant simplification of calculations appears if instead of (\ref{Vef}) 
we use the model interaction of the form (similar simplification was first
used in Ref.\cite{Kam}):
\begin{equation}
V_{eff}({\bf q})=W^2\frac{2\xi^{-1}}{\xi^{-2}+(q_x-Q_x)^2}
\frac{2\xi^{-1}}{\xi^{-2}+(q_y-Q_y)^2}
\label{Veff}
\end{equation}
where $W^2=\tilde W^2/4$. In fact (\ref{Veff}) is qualitatively quite similar
to (\ref{Vef}) and differs from it only slightly quantitatively in most
interesting region of $|{\bf q-Q}|<\xi^{-1}$.
 Analogous but a little bit
different form of effective interaction was used in Ref.\cite{Sch}. In this way
we achieve effectively one - dimensional situation.

\begin{figure}
\epsfxsize=14cm
\epsfysize=9cm
\epsfbox{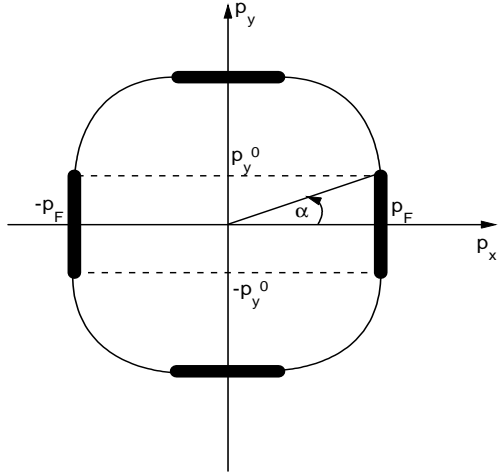}
\caption{Fermi surface in the model of ``hot patches'', which are shown by
thick lines (e.g. of the width of $\sim\xi^{-1}$). The angle $\alpha$ 
determines the size of a ``hot patch'' \cite{PS}, $\alpha=\pi/4$  corresponds
to the square Fermi surface.}
\label{hpatches}
\end{figure}

Scattering by AFM fluctuations in HTSC -- cuprates is not always most
intensive at vector ${\bf Q}=(\pi/a,\pi/a)$ commensurate with the inverse
lattice vectors, in general case vector ${\bf Q}$
 may correspond to
incommensurate scattering. Keeping in mind the observed topology of the Fermi
surface with flat parts (patches) as shown in Fig.\ref{FSBISCO}, we may propose
another model for scattering by fluctuations of AFM short - range order
\footnote{As in the following we always neglect, for simplicity, the spin 
structure of interaction, our analysis is, strictly speaking, applicable to
the case of electron interaction with fluctuations of charge density wave (CDW) 
short - range order. However, this simplification is unimportant if we are 
interested only in qualitative results for pseudogap renormalization of
electronic spectrum. Note that interaction with fluctuations of CDW type was
proposed  as an alternative mechanism of pseudogap anomalies of cuprates in a
number of papers, see e.g.\cite{DiCas}.}, which we shall call the ``hot patches'' model
\cite{PS}. Assume that the Fermi surface of our two - dimensional system of
electrons has the form, shown in Fig.\ref{hpatches}. 
The size of ``hot patch''
is determined by the angular parameter $\alpha$.
 It is well known that the
presence of flat parts on the Fermi surface leads to instabilities either
of charge density wave (CDW) or spin density wave (SDW) type, with the 
establishment of the appropriate long - range order and opening of
(dielectric) energy gap on these parts of the Fermi surface.
Here we are interested in situation typical of fluctuation region of the
appropriate phase transition, when there is still no long - range order present
\footnote{Assumption about the existence of flat parts on the Fermi surface
is in fact not crucial for our model, but simplifies calculations considerably,
which can always be done in more realistic ``hot spot'' model.}. 
Similar model of the Fermi surface was introduced a number of years ago as
a possible model for HTSC -- cuprates in Refs.\cite{Vir,Ruv,Dz},
 where rather
detailed analysis of microscopic criteria for the appearance of 
antiferromagnetic and superconducting phases on the phase diagram was
performed.
Fluctuations of short - range order are assumed to be static and Gaussian
\footnote{This assumption, strictly speaking, is also applicable only for
high enough temperatures.} and the appropriate correlation function is
defined as (similar to 
\cite{Kam}):  
\begin{equation}
S({\bf q})=\frac{1}{\pi^2}\frac{\xi^{-1}}{(q_x-Q_x)^2+\xi^{-2}}
\frac{\xi^{-1}}{(q_y-Q_y)^2+\xi^{-2}}
\label{fluct}
\end{equation}
Here $\xi$ -- is again the correlation length of fluctuations, while the
scattering vector is taken as $Q_x=\pm 2k_F$,\ $Q_y=0$ or 
$Q_y=\pm 2k_F$,\ $Q_x=0$.
 It is assumed that only electrons from flat (``hot'')
parts of the Fermi surface shown in Fig.\ref{hpatches} interact with these 
fluctuations and this scattering is in fact one - dimensional.
For the special case of $\alpha=\pi/4$ we obtain square Fermi surface and the
problem becomes strictly one - dimensional. For $\alpha<\pi/4$ there are also
``cold'' parts present at the Fermi surface and the scattering there is 
either absent or weak enough. Effective interaction of electrons from
``hot'' patches with fluctuations will be descibed by 
$(2\pi)^2W^2S({\bf q})$, 
where $W$ is the parameter with dimension of energy, defining again the
energy scale (width) 
of the pseudogap.
\footnote{It can be said that we introduce an effective interaction
``constant'' with fluctuations of the following form:
$W_{\bf p}=W[\theta(p_x^0-p_x)\theta(p_x^0+p_x)+\theta(p_y^0-p_y)
\theta(p_y^0+p_y)]$ (see Fig.\ref{hpatches}).}.  
The choice of scattering vector ${\bf Q}=(\pm 
2k_F,0)$ or 
${\bf Q}=(0,\pm 2k_F)$  assumes the picture of incommensurate
fluctuations as the Fermi momentum $p_F=\hbar k_F$ is in general not connected
with the inverse lattice period. Commensurate case can also be studied within
this model \cite{PS}.

The basic idea of models under discussion is the presence of strong
scattering on fluctuations of short - range order, which, according to
(\ref{V}), (\ref{Veff}) or (\ref{fluct}), is effective in a limited region of
inverse space with the width of the order of $\xi^{-1}$
 around the ``hot''
spots or patches, and which leads to pseudogap -- like renormalization of
electronic spectrum in these regions\footnote{Analogous situation is
realized in liquid metals or semiconductors, where the information about the
lost crystal structure is partially conserved in the so called structure
factor with characteristic maximum in momentum space, being the crucial element
of electronic theory of these systems according to Ziman and Edwards 
\cite{Zimn}. However, in three - dimensional isotropic liquids the angular
averaging leads, in general, to rather strong suppression of pseudogap
effects.}. 

\begin{figure}
\epsfxsize=12cm
\epsfysize=11cm
\epsfbox{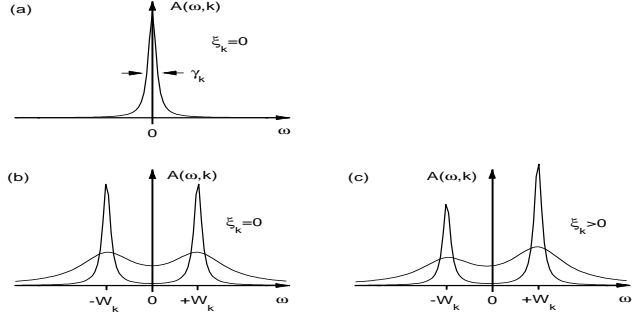}
\caption{Qualitative evolution of the spectral density.
(a) -- normal metal (Fermi - liquid), $\xi_{\bf k}=0$ -- at the Fermi surface.
(b) -- two narrow peaks corresponding to Bogoliubov's quasiparticles in the
system with dielectric gap $W_{\bf k}$ (CDW or SDW long - range order). 
Smeared maxima -- system without long - range order (pseudogap behavior),
$\xi_{\bf k}=0$ -- at the Fermi surface.
(c) The same as in (b), but for $\xi_{\bf k}>0$, i.e. above the Fermi surface.
In this case a characteristic asymmetry of the spectral density appear.}
\label{sdqual}
\end{figure}

Let us present a simple qualitative analysis of possible changes of
single - electron spectral density (\ref{sdens}). In the standard Fermi - liquid
theory \cite{AGD} the single - electron Green's function for the metal can be
written as:
\begin{equation}
G(\omega,{\bf k})=\frac{Z_{\bf k}}{\omega-\xi_{\bf k}-i\gamma_{\bf k}}+
G_{incoh}
\label{FLGr}
\end{equation}
where $\xi_{\bf k}=\varepsilon_{\bf k}-\mu$ -- is the quasiparticle energy
with respect to the Fermi level (chemical potential) $\mu$, $\gamma_{\bf k}$ -- 
its damping. The residue in the pole $0<Z_{\bf k}<1$, $G_{incoh}$ -- 
non - singular contribution of many - particle excitations. Then the
appropriate spectral density is:
\begin{equation}
A(\omega,{\bf k})=\frac{1}{\pi}Z_{\bf k}\frac{\gamma_{\bf k}}
{(\omega-\xi_{\bf k})^2+\gamma^2_{\bf k}}+...
\label{FLsdens}
\end{equation}
where dots denote more or less regular contribution due to
$G_{incoh}$, while the quasiparticle spectrum is reflected in a narrow
(as damping $\gamma_{\bf k}$ is small close to the Fermi level) Lorentzian peak. 
This 
situation is illustrated in Fig.\ref{sdqual}(a).

If the long - range order of SDW or CDW type appears the spectrum of
elementary excitations acquires the (dielectric) gap $W_{\bf k}$ 
(dependence on ${\bf k}$ stresses the possibility of gap opening on some
part of the Fermi surface only) and single - electron Green's function
has the form of Gorkov's function \cite{SWZ,Izy}, where we can add some damping
$\Gamma_{\bf k}$:
\begin{equation}
G(\omega,{\bf k})=\frac{u^2_{\bf k}}{\omega-E_{\bf k}+i\Gamma_{\bf k}}
+\frac{v^2_{\bf k}}{\omega+E_{\bf k}-i\Gamma_{\bf k}}
\label{GorkGF}
\end{equation}
where the spectrum of elementary excitations:
\begin{equation}
E_{\bf k}=\sqrt{\xi^2_{\bf k}+W^2_{\bf k}}
\label{EkGrk}
\end{equation}
and we introduced Bogoliubov's coefficients:
\begin{eqnarray}
u^2_{\bf k}=\frac{1}{2}\left(1+\frac{\xi_{\bf k}}{E_{\bf k}}\right) \\
v^2_{\bf k}=\frac{1}{2}\left(1-\frac{\xi_{\bf k}}{E_{\bf k}}\right)
\label{u-v}
\end{eqnarray}
Then the spectral density:
\begin{equation}
A(\omega{\bf k})=
\frac{u^2_{\bf k}}{\pi}\frac{\Gamma_{\bf k}}
{(\omega-E_{\bf k})^2+\Gamma^2_{\bf k}}+
\frac{v^2_{\bf k}}{\pi}\frac{\Gamma_{\bf k}}
{(\omega+E_{\bf k})^2+\Gamma^2_{\bf k}}+...
\label{GorkSDens}
\end{equation}
where now {\em two} peaks are present (both narrow if $\Gamma_{\bf k}$ is small),
corresponding to Bogoliubov's quasiparticles.

Let us now consider the situation when SDW or CDW long - range order is
absent, but there is a strong scattering on fluctuations of short - range 
order with appropriate wave vectors
 (cf. (\ref{V}),(\ref{Veff}) or (\ref{fluct})). 
Then it is not difficult to imagine that in the spectral density remains some
``reminiscence'' (or ``anticipation'') of dielectric energy gap opening (in
the same region of momentum space) in the form of characteristic smeared
two - peak structure, as shown in Fig.\ref{sdqual}
\footnote{In the following the term ``non - Fermi liquid behavior'' will be
used only in this, rather limited, sense.}. 
The width of both maxima, 
naturally, is determined by some parameter like $v_F/\xi$, 
i.e. by inverse
time of flight of an electron through the region of the typical size of
$\xi$,
 where effectively persists ``dielectric'' ordering.
It is seen that this qualitative picture is quite similar to ARPES data
shown in Fig.\ref{Isym}. Below we shall see that more rigorous analysis leads
just to the same conclusions. Further parts of the review will be dealing with
justification of this qualitative picture and some of its consequences.

\subsection {Recurrence Relations for the Green's Functions.}

The model of the pseudogap state which will be discussed below is, in fact,
a two - dimensional generalization of the model proposed many years ago for
pseudogap formation in fluctuation region of Peierls (CDW) structural
transition in one - dimensional systems \cite{LRA,C1,C2,C79}
\footnote {In Ref.\cite{C1} this model
was proposed for the explanation of pseudogap 
in liquid semiconductors}. 
In Refs.\cite{C1,C2} we obtained an exact solution of this model in the limit
of very large correlation lengths of fluctuations of short - range order
$\xi\to\infty$, while in Ref.\cite{C79} we proposed the appropriate 
generalization for the case of finite $\xi$. The main advantage of these
results, in comparison with Ref.\cite{LRA},
 was the complete account of all
Feynman diagrams of perturbation series for electron interaction with
fluctuations of short - range order. Appropriate two - dimensional
generalization of this approach was made in Refs.\cite{Sch,KS} for ``hot spots''
model, while for ``hot patches'' model it can be done directly due to
effective one - dimensionality of this model.

Contribution of an arbitrary diagram for electron self - energy of single -
particle Green's function in the $N$-th order of perturbation theory over
interaction with fluctuations, given by (\ref{Veff}) or (\ref{fluct}),
 can be
approximately written in the following form \cite{KS}:
\begin{equation}
\Sigma^{(N)}(\varepsilon_n{\bf p})=W^{2N}\prod_{j=1}^{2N-1}
\frac{1}{i\varepsilon_n-\xi_{j}+in_jv_j\kappa}
\label{Ansatz}
\end{equation}
where for the ``hot spots'' model
$\xi_j=\xi_{\bf p+Q}$ and $v_j=|v_{\bf p+Q}^{x}|+|v_{\bf p+Q}^{y}|$ for odd
$j$ and $\xi_j=\xi_{\bf p}$ and $v_{j}=|v_{\bf p}^x|+|v_{\bf p}^{y}|$
for even $j$. In the  ``hot patches'' model $\xi_j=(-1)^j\xi_{\bf p}$,
$v_j=v_F$. In (\ref{Ansatz}) we have introduced an inverse correlation length 
$\kappa=\xi^{-1}$, $n_j$ -- is the number of interaction lines, surrounding 
$j$-th (from the beginning) Green's function in a given diagram,
$\varepsilon_n=
(2n+1)\pi T$ and, for definiteness, $\varepsilon_n>0$.

As a matter of fact, expression (\ref{Ansatz}) constitutes rather
successful {\em Ansatz}, which allows to calculate the contribution of an
arbitrary diagram in any order. In this case any diagram with intersections of
interaction lines is actually equal to some specific diagram of the same order
without intersecting interaction lines. Thus, in fact we may consider only
diagrams without intersections of interaction lines, taking those with
intersections into account by additional combinatorial factors, associated
with interaction lines. This approach was first used (for another problem)
by Elyutin \cite{Ely} and applied to one - dimensional model of the pseudogap
state in Ref.\cite{C79}.
 Actually, {\em Ansatz} (\ref{Ansatz}) is not exact
\cite{Tchern}, 
but it may be justified in two - dimensional case for
certain topologies of the Fermi surface, such that electron velocity components
in ``hot spots'' connected by vector ${\bf Q}$ have the same sign, so that
$v_{\bf p}^xv_{\bf p+Q}^x>0$ and $v_{\bf p}^yv_{\bf p+Q}^y>0$ \cite{KS}. 
This condition is not satisfied in one - dimensional case and also for
topologies of Fermi surfaces typical for HTSC -- cuprates, as shown in
Fig.\ref{hspots}.
 However, in Ref.\cite{KS} it was shown that even in these
cases the use of (\ref{Ansatz})
 gives quite satisfactory results and
reproduce the exactly known limits of $\xi\to\infty$ and $\xi\to 0$ \footnote{
Even in most unfavorable case of one - dimensional model {\em Ansatz}
(\ref{Ansatz}) gives very good quantitative approximation, e.g. for the
density of states, which is seen from the direct comparison \cite{CKop}
 of
the results of \cite{C79} for incommensurate case with the results of an
exact numerical simulation of Refs.\cite{Kop,MilMoni}. In commensurate case
the use of (\ref{Ansatz}) in one - dimensional model does not reproduce only
the so called Dyson singularity of the density of states at the center of
pseudogap \cite{Kop,MilMoni}, which is apparently absent in two dimensions.}.

As a result we obtain the following recursion relation for one - electron
Green's function (continuous chain approximation \cite{C79}):
\begin{equation}
G^{-1}(\varepsilon_n\xi_{\bf p})=G_{0}^{-1}(\varepsilon_n\xi_{\bf p})-
\Sigma_{1}(\varepsilon_n\xi_{\bf p})
\label{G}
\end{equation}
\begin{equation}
\Sigma_{k}(\varepsilon_n\xi_{\bf p})=W^2\frac{v(k)}
{i\varepsilon_n-\xi_k+ikv_k\kappa-\Sigma_{k+1}(\varepsilon_n\xi_{\bf p})}
\label{rec}
\end{equation}
Combinatorial factor:
\begin{equation}
v(k)=k
\label{vcomm}
\end{equation}
corresponds to the case of commensurate fluctuations with
${\bf Q}=(\pm\pi/a,\pm\pi/a)$ \cite{C79}. There is no problem to analyze also
the case of incommensurate fluctuations when $\bf Q$ is not related to the
inverse lattice period. In this case diagrams with interaction lines
surrounding odd number of vertices are much smaller than those with
interaction lines surrounding the even number of vertices. So we have to take 
into account these latter diagrams \cite{C1,C2,C79}.
 In these case the
recursion relation (\ref{rec}) is still valid, but diagram combinatorics
change, so that combinatorial factor
s $v(k)$ are:
\cite{C79}:
\begin{equation}
v(k)=\left\{\begin{array}{cc}
\frac{k+1}{2} & \mbox{for odd $k$} \\
\frac{k}{2} & \mbox{for even $k$}
\end{array} \right.
\label{vincomm}
\end{equation}
In Ref.\cite{Sch} the spin structure of interaction was taken into account 
in the framework of the model of ``nearly antiferromagnetic'' Fermi - liquid 
(spin - fermion model). This leads to more complicated diagram combinatorics
for commensurate case with 
${\bf Q}=(\pm\pi/a,\pm\pi/a)$. Spin - conserving
scattering leads formally to commensurate combinatorics, while spin - flip
scattering is described by diagrams as for 
incommensurate case
(``charged'' random field in terms of Ref.\cite{Sch}). As a result, the
recursion relation for the Green's function is still given by (\ref{rec}), 
but the combinatorial factor $v(k)$ becomes \cite{Sch}:
\begin{equation}
v(k)=\left\{\begin{array}{cc}
\frac{k+2}{3} & \mbox{for odd $k$} \\
\frac{k}{3} & \mbox{for even $k$}
\end{array} \right.
\label{vspin}
\end{equation}

\subsection {Spectral Density and the Density of States.}

\begin{figure}
\epsfxsize=9cm
\epsfysize=9cm
\epsfbox{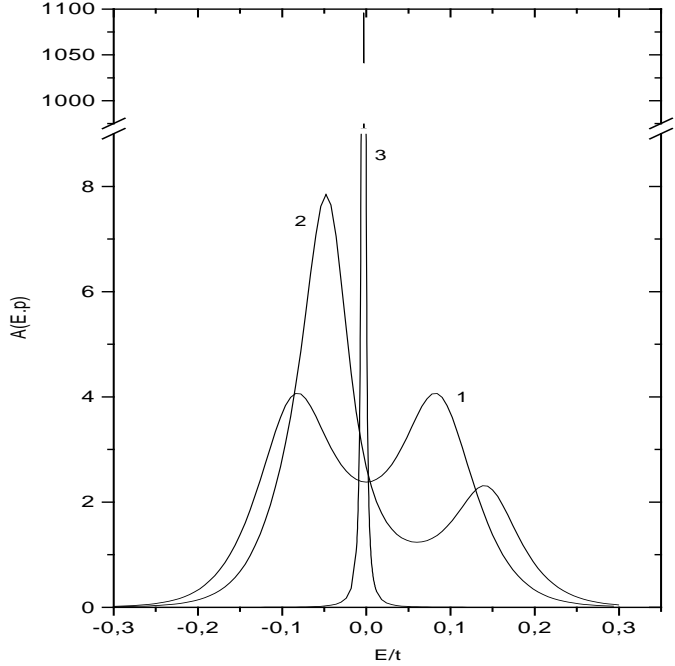}
\caption{Spectral density in ``hot spots'' model, incommensurate case,
$\kappa a=0.01$ \cite{KS}. (1) -- at the ``hot spot'' 
$p_xa/\pi=0.142, p_ya/\pi=0.587$. (2) -- close to the ``hot spot'' for 
$p_xa/\pi=0.145, p_ya/\pi=0.843$. (3) -- far from the ``hot spot'' for
$p_xa/\pi=p_y/\pi=0.375$.}
\label{sdnshs}
\end{figure}

\begin{figure}
\epsfxsize=9cm
\epsfysize=9cm
\epsfbox{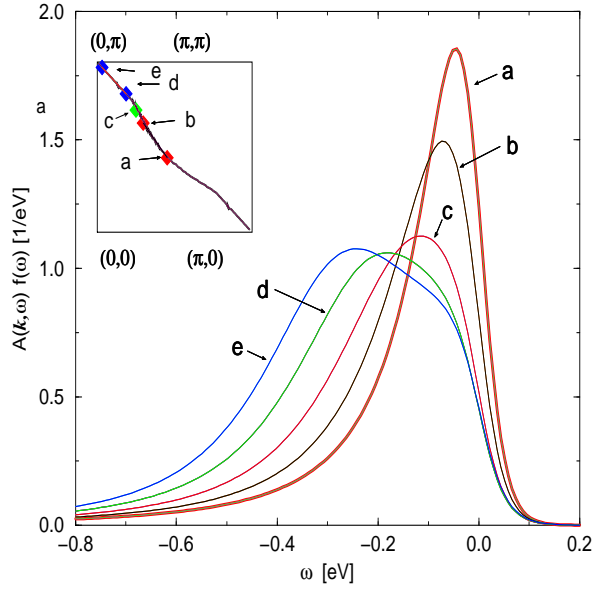}
\caption{Product of spectral density and Fermi distribution function at
different points of the Fermi surface, shown at the insert.
Spin - fermion model, correlation length $\xi=3a$ \cite{Sch}.}
\label{fspdns}
\end{figure}

Recurrence relations (\ref{G}) and (\ref{rec}) are quite convenient for
numerical calculations. Rather detailed studies of
spectral density, ARPES characteristics and density of states for
different variants of ``hot spots'' model were performed in Refs.\cite{Sch,KS}.
As a typical example in Fig.{\ref{sdnshs}} we present the results of \cite{KS}
for the spectral density for incommensurate case.
We can see that the spectral density close to the ``hot spot'' has the expected
non - Fermi liquid like form, quasiparticles can not be defined. Far from the
``hot spot'' the spectral density reduces to a sharp peak, corresponding to
well defined quasiparticles (Fermi - liquid). In Fig.\ref{fspdns} from
Ref.\cite{Sch} we show the product of Fermi distribution and spectral density
for different points on ``renormalized'' Fermi surface, defined by the
equation $\varepsilon_{\bf k}-Re\Sigma(\omega=0{\bf k})=0$,
 where
the ``bare'' spectrum $\varepsilon_{\bf k}$ was taken as in (\ref{tbspectr})
with $t=-0.25eV, t'=-0.35t$ for hole concentration $n_h=0.16$, coupling
constant in (\ref{V}) $g=0.8eV$ and correlation length $\xi=3a$ 
(commensurate case of spin - fermion model).
The qualitative agreement with ARPES data discussed above is clearly seen, as
well as qualitatively different behavior close and far from the ``hot spot''.
\begin{figure}
\epsfxsize=9cm
\epsfysize=9cm
\epsfbox{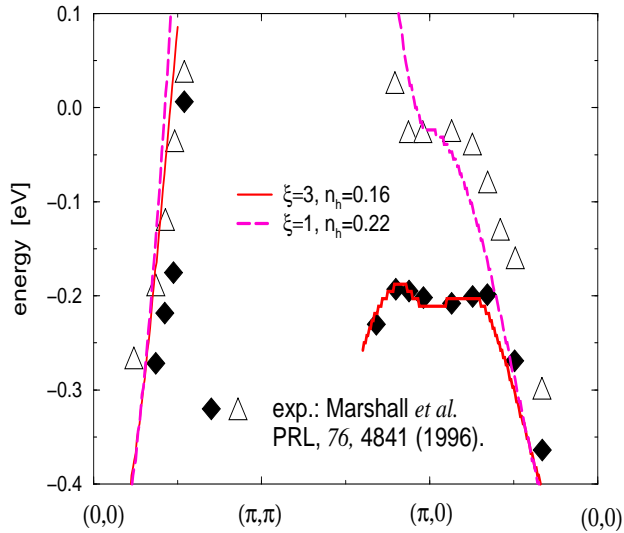}
\caption{Position of the maximum of spectral density for two values of
correlation length $\xi$ and hole concentration, calculated for spin -
fermion model \cite{Sch} in comparison with ARPES data of Ref.\cite{Marsh} 
for $Bi_2Sr_2Ca_{1-x}Dy_{x}Cu_2O_{8+\delta}$ 
with $x=1$ (triangles) and $x=0.175$ (diamonds).}
\label{locmax}
\end{figure}
In Fig.\ref{locmax} from \cite{Sch} we show calculated (for spin - fermion
model with interaction (\ref{V}), static limit) positions of the maximum of
$A(\omega{\bf k})$ for two different concentrations of holes and appropriate
experimental ARPES - data of
 Ref.\cite{Marsh} for
$Bi_2Sr_2Ca_{1-x}Dy_{x}Cu_2O_{8+\delta}$. These maxima positions in 
$(\omega,{\bf k})$ plane, obtained from ARPES, in fact determine (for Fermi -
liquid like system) the experimental dispersion curves of quasiparticles 
(Cf. Fig.\ref{sdqual}(a)). For overdoped system the values of
$n_h=0.22$ and $\xi=a$ were assumed. The results show rather well defined
branches of the spectrum both in the direction of diagonal of the
Brillouin zone, as well as in the 
 $(0,0)-(\pi,0)$ direction. For underdoped
system the values of $n_h=0.16$ and $\xi=3a$ were used. In this case, in
diagonal direction we can again see the crossing of the spectrum with the Fermi
level, while close to the ``hot spots'' in the vicinity of $(\pi,0)$ the
smeared maximum of the spectral density remains approximately $200meV$ 
lower than the Fermi level (pseudogap behavior). In general, the agreement
of theoretical model and experiment is rather encouraging.

Consider now the single - electron density of states:
\begin{equation}
N(E)=\sum_{\bf p}A(E,{\bf p})=-\frac{1}{\pi}\sum_{\bf p}ImG^R(E{\bf p})
\label{dos}
\end{equation}
defined by the integral of spectral density $A(E{\bf p})$ over Brillouin zone.
Detailed calculations of the density of states in the ``hot spots'' model were
performed in Ref.\cite{KS}. For typical value of $t'/t=-0.4$
  rather smooth
suppression of the density of states (pseudogap) was obtained. This drop in the
density of states is relatively weakly dependent on the value of
correlation length $\xi$. 
At the same time, for $t'/t=-0.6$ (non characteristic
for HTSC -- cuprates) the ``hot spots'' on the Fermi surface still exist, but
pseudogap in the density of states is practically invisible. 
Only the strong
smearing of Van - Hove singularity, existing for the system without 
fluctuations, is clearly observed.

In ``hot patches'' model the use of (\ref{G}) and (\ref{rec}) 
leads to the
spectral density on these patches qualitatively similar to that shown in
Fig.\ref{sdnshs} \cite{C91a,McK}. 
On ``cold''parts of the Fermi surface the
spectral density reduces to $\delta$-function, typical for Fermi - liquid
(gas) (Cf. Fig.\ref{sdqual} (a)). 
The density of states can be written as:
\begin{equation}
N(E)=-\frac{1}{\pi}N_0(0)\int_{0}^{2\pi}\frac{d\phi}{2\pi}
\int \limits_{-\infty}^{\infty}d{\xi_p} Im G^R(E\xi_p)=
\frac{4\alpha}{\pi} N_W(E)+(1-\frac{4\alpha}{\pi})N_0(0)
\label{NN}
\end{equation}
where $N_0(0)$ - is the density of states of free electrons at the Fermi level,
while $N_W(E)$ - is the density of states of one - dimensional model
(square Fermi surface), obtained previously in Refs.\cite{C1,C2,C79}.

Let us quote some more detailed results for (artificial enough) limit of
very large correlation lengths $\xi\to\infty$ (``toy'' -- model). 
In this case the whole perturbation series for the electron scattered by 
fluctuations is easily summed \cite{C1,C2} and an exact analytic solution for 
the Green' function is obtained in the following form \cite{PS}:  
\begin{equation} 
G(\epsilon_n,p)=\int \limits_0^\infty 
dD {\cal P}(D) 
\frac{i\epsilon_n+\xi_p}{(i\epsilon_n)^2-\xi_p^2- D(\phi)^2},
\label{fgrina}
\end{equation}
where $D(\phi)$ is defined for $0\leq\phi\leq\frac{\pi}{2}$ as:
\begin{equation}
D(\phi)=\left\{
\begin{array}{ll}
D & ,0\leq\phi\leq\alpha,\>\frac{\pi}{2}-\alpha\leq\phi\leq\frac{\pi}{2} \\
0 & ,\alpha\leq\phi\leq\frac{\pi}{2}-\alpha
\end{array}
\right.
\label{w}
\end{equation}
For other values of $\phi$ the value of $D(\phi)$ is determined by obvious
symmetry analogously to (\ref{w}). 

The amplitude of dielectric gap $D$ is random and distributed according to
Rayleigh distribution \cite{C79} (its phase is also random and distributed
homogeneously on the interval $(0,2\pi)$):
\begin{equation}
{\cal P}(D)=\frac{2D}{W^2}exp\left(-\frac{D^2}{W^2}\right)
\label{Rayl}
\end{equation}
Thus, on the ``hot patches'' the Green's function has the form of ``normal''
Gorkov's function, averaged over spatially homogeneous fluctuations of
dielectric gap $D$, distributed according to \ref{Rayl}. ``Anomalous''
Gorkov's function is, however, zero due to random phases of dielectric gap
$D$, which corresponds to the absence of the long - range order\footnote{
Note that averages of pairs of anomalous functions are different from zero and
contribute to the appropriate exact expression for two - particle Green's
function \cite{C1,C2}}.

For finite correlation lengths $\xi$ the amplitude of our one - dimensional
random ``periodic'' field is approximately constant on the lengths of the 
order of $\xi$, while its values in different regions with sizes of the order
of $\sim\xi$ are random. The crossover from one region with approximately
periodic field to the other also takes place at lengths of the order of
$\xi$. The electron is effectively scattered only during this transition from 
one region to another, which takes a time of the order of $\sim\xi/v_F$, 
leading to effective damping of the order of $\sim v_F/\xi$.
 Interesting data
on numerical modeling of our random field, for one - dimensional case, can be
found in Ref.\cite{Tchern}.

Outside ``hot patches'' (the second inequality in (\ref{w})) the Green's
function (\ref{fgrina}) just coincides with that of free electrons.

Density of states, corresponding to (\ref{fgrina}), has the form (\ref{NN}), 
where
\begin{equation}
\frac{N_W(E)}{N_0(0)}=\left|\frac{E}{W}\right|
\int\limits_0^\frac{E^2}{W^2}d{\zeta}\frac{e^{-\zeta}}
{\sqrt{\frac{E^2}{W^2}-\zeta}}=
2\left|\frac{E}{W}\right|
exp(-\frac{E^2}{W^2})Erfi(\frac{E}{W})
\label{NW}
\end{equation}
where $Erfi(x)$ -- is the probability integral of imaginary argument.
\begin{figure}
\epsfxsize=9cm
\epsfysize=11cm
\epsfbox{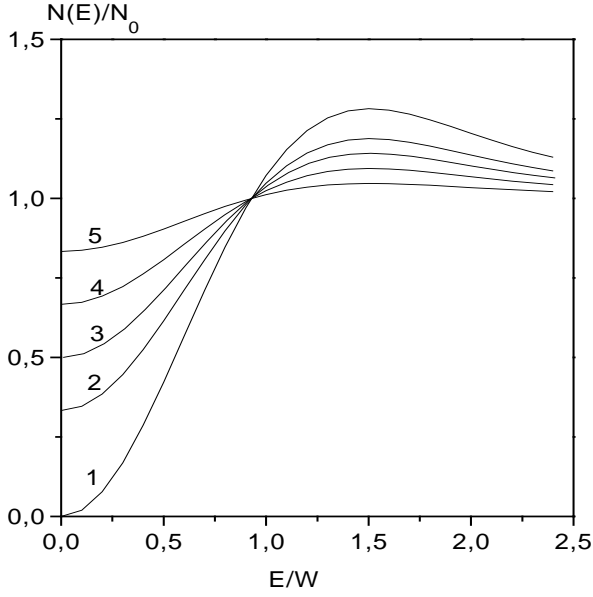}
\caption{Density of states for ``hot patches'' of different sizes
\cite{PS}: (1) -- $\alpha=\pi/4$;\ (2) -- $\alpha=\pi/6$;\ 
(3) -- $\alpha=\pi/8$;\ 
(4) -- $\alpha=\pi/12$;\ (5) -- $\alpha=\pi/24$.
}
\label{Nalp}
\end{figure}
In Fig.\ref{Nalp} we show density of states of our model for different values
of parameter $\alpha$, i.e. for  ``hot patches'' of different sizes. It is seen
that the pseudogap in the density of states is rather fast smeared 
(or filled in) as ``hot patches'' become smaller. In some sense the effect of
diminishing  $\alpha$ is analogous to the effect of diminishing correlation
length $\xi$ \cite{C79}, so that the approximation of  $\xi\rightarrow\infty$, 
is probably not so very strong limitation of this model. Finite values of
$\xi$ can be easily accounted using (\ref{G}) and (\ref{rec}), leading to
additional smearing of the pseudogap for smaller $\xi$. Note the general
qualitative agreement of the form of the pseudogap in this model with those
observed in tunneling experiments, e.g. shown in Figs.\ref{tunn},\ref{tunnH}.

\subsection {Two-Particle Green' Function and Optical Conductivity.}

Remarkably, for these models it is possible to calculate explicitly also
two - particle Green's functions for the electron in the random field of
short - range order fluctuations \cite{C91,C91a} (Cf. also Ref.
\cite{Sch}), 
taking into account {\em all} Feynman diagrams of perturbation series.
As calculations for ``hot spots'' model require rather large numerical work
due to the use of ``realistic'' spectrum of current carriers (\ref{tbspectr}), 
we shall limit ourselves to the simplified analysis within ``hot patches''
model \cite{C99}. 

On ``cold'' parts of the Fermi surface we shall assume the existence of some
small static scattering of an arbitrary nature, describing it by
phenomenological scattering frequency  $\gamma$, assuming, in most cases, that
$\gamma\ll W$,
 so that we may neglect this scattering on ``hot patches''.
Accordingly, on ``cold'' parts the electronic spectrum will be described by
by usual expressions for the Green's functions for the system with weak
(static) scattering.

Consider first the limit of $\xi\rightarrow \infty$
, when the single - electron
Green's function has the form (\ref{fgrina}) and two - particle Green's 
function can also be found exactly using methods of Refs.\cite{C1,C2}.   

Conductivity in this model always consists of additive contributions from
``hot'' and  ``cold'' parts of the Fermi surface, analogously to (\ref{NN}).
In particular, for its real part in the limit of $\xi\to\infty$ we have:
\begin{equation}
Re\sigma(\omega)=\frac{4\alpha}{\pi}Re\sigma_{W}(\omega)
+(1-\frac{4\alpha}{\pi})Re\sigma_D(\omega)
\label{Resigma}
\end{equation}
where, with the account of the results of Refs.\cite{C1,C2}:
\begin{equation}
Re\sigma_{W}(\omega)=\frac{\omega_p^2}{4}\frac{W}
{\omega^2}\int \limits_0^{\omega^2/4W^2}
d{\zeta} exp(-\zeta)\frac{\zeta}{\sqrt{\omega^2/4W^2-\zeta}}
\label{Dsig}
\end{equation}
where $\omega_p$ -- is plasma frequency and
\begin{equation}
Re\sigma_D(\omega)=\frac{\omega_p^2}{4\pi}\frac{\gamma}{\omega^2+\gamma^2}
\label{Drude}
\end{equation}
-- the usual Drude - like contribution of ``cold'' parts.

Even in this simplest approximation the dependence of $Re\sigma$ on $\omega$
is quite similar to observed in the experiments \cite{PBT,STP,B1,B2}. It is
characterized by narrow  ``Drude'' peak at small frequencies and smooth
pseudogap absorption maximum  at $\omega\sim 2W$. 
As scattering frequency
$\gamma$
 on ``cold'' parts grows, characteristic ``Drude'' peak at small
frequencies is suppressed.

More realistic case of finite correlation lengths of fluctuations of
``antiferromagnetic'' short range order  $\xi$ in (\ref{fluct}) can be
analyzed using (\ref{G}), (\ref{rec}). 
Vertex part
$J^{RA}(\epsilon,\xi_{\bf p};\epsilon+\omega,\xi_{\bf p+q})$, 
defining density - density response function (two - particle Green' function)
on ``hot patches''an be determined from the following recursion procedure
(details can be found in Refs.\cite{C91,C91a} and also in 
\cite{Sch}), which takes into account all diagrams of perturbation series for
interaction with fluctuations:
\begin{eqnarray} 
\label{vertex}
J^{RA}_{k-1}(\epsilon,\xi_{\bf p};\epsilon+\omega,\xi_{\bf p+q})=\\ \nonumber
=e+W^2v(k)G^A_k(\epsilon,\xi_{\bf p})G^R_k(\epsilon+\omega,\xi_{\bf 
p+q})J^{RA}_{k}(\epsilon,\xi_{\bf p};\epsilon+\omega,\xi_{\bf p+q})\times \\
\nonumber
\times\Biggl\{1+\frac{2iv_F \kappa k}{\omega-(-1)^k v_Fq+v(k+1)W^2
[G^A_{k+1}(\epsilon,\xi_{\bf p})-G^R_{k+1}(\epsilon+\omega,\xi_{\bf p+q})]} 
\Biggr\}
\end{eqnarray}
where $e$ -- is electron charge, $R(A)$ denotes retarded (advanced)
Green's function, and the vertex part of interest to us is defined as
$k=0$  term of this recursion procedure. Contribution of ``hot patches'' to
conductivity 
$Re\sigma_{W}(\omega)$ in (\ref{Resigma}) is calculated as in 
Refs.\cite{C91,C91a}, while $Re\sigma_D(\omega)$ is again determined
by (\ref{Drude}). 

\begin{figure}
\epsfxsize=8cm
\epsfysize=10cm
\epsfbox{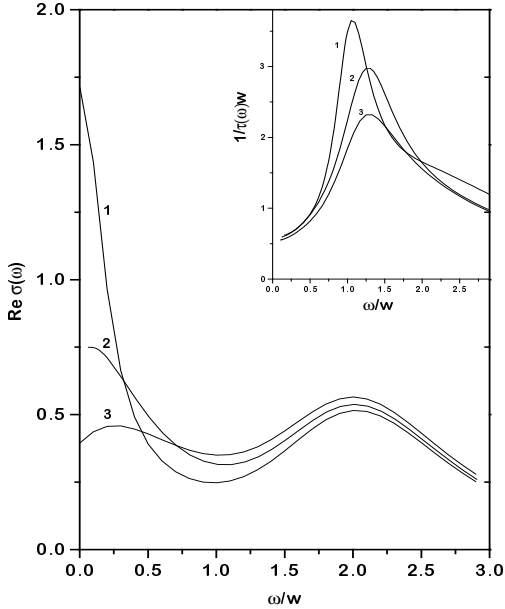}
\caption{Real part of conductivity as a function of frequency of external field,
for different values of $\gamma$
 and fixed correlation length $v_F\kappa=0.5W$.
Conductivity is in units of $\omega_p^2/4\pi W$.\ 
(1)--$\gamma/W=0.2$;\ 
(2)--$\gamma/W=0.5$;\ 
(3)--$\gamma/W=1.0$. The size of ``hot patches'' $\alpha=\pi/6$. \cite{C99}.
At the insert -- calculated dependence of generalized scattering frequency
$1/\tau$
 (in units of $W$), as a function of frequency for different values
of correlation length and fixed $\gamma=0.2W$.\
(1)--$v_F\kappa=0.1$;\
(2)--$v_F\kappa=0.5$;\
(3)--$v_F\kappa=1.0$ \cite{Str}}
\label{condopt}
\end{figure}

Typical results of such calculations are shown in Fig.\ref{condopt}
for the case of incommensurate fluctuations. 
The use of combinatorics of spin - fermion model leads to rather insignificant
quantitative changes. The general qualitative picture remains the same also for
commensurate case \cite{C99}. Conductivity is always characterized by
narrow enough ``Drude'' peak in the region of small frequencies
$\omega < \gamma$, originating from ``cold'' parts of the Fermi surface and
a smooth maximum in the region of $\omega\sim 2W$, 
corresponding to the
absorption through the pseudogap opened at the ``hot patches'.'
``Drude'' peak is rapidly suppressed as $\gamma$ grows. Dependence on
correlation length of fluctuations $\xi=\kappa^{-1}$ in most interesting 
parameter region is weak enough. 
This qualitative picture is very similar
to experimental dependencies observed for many HTSC -- systems 
\cite{PBT,STP,B1,B2}, 
typical data were shown above in Fig.\ref{optcond}. 
In this model we can also calculate the imaginary part of conductivity and
the parameters of the generalized Drude model (\ref{sgdr}), (\ref{mgdr}) 
\cite{Str}.
 Calculated $1/\tau(\omega)$ is shown at the insert in
Fig.\ref{condopt} 
and demonstrates pseudogap behavior in the frequency
region of $\omega<W$, quite analogous to that shown in 
Fig.\ref{genDr}\footnote{For $\omega>2W$ the linear growth of
$1/\tau(\omega)$ with frequency is not reproduced, but it can be easily
obtained if we use a phenomenological dependence 
$\gamma({\omega})=\gamma_0+a\omega$ in the spirit of ``marginal'' Fermi -
liquid.}. 
Apparently it is not difficult to fit experimental data to our
dependences using the known values of $\omega_p\sim 1.5-2.5eV$ 
and $2W\sim 0.1eV$, as well as ``experimental
'' values of $\gamma$, 
determined
from the width of ``Drude'' peak, and varying ``free'' parameters
$\alpha$ (the size of ``hot patches'')
 and $\xi$ (the value of this can also
be taken from other experiments \cite{Sch}).

\section {Superconductivity in the Pseudogap State.}

\subsection {Gorkov's Equations.}

At present there is no agreement on microscopic mechanism of Cooper pairing
in HTSC - cuprates. The only well established fact is the anisotropic nature
of pairing with $d$-wave symmetry \cite{Legg,TsuK}, though even here some
alternative points of view exist. Apparently, the majority of researchers
are inclined to believe in some kind pairing mechanism due to an exchange by
spin (antiferromagnetic) excitations. A recent review of these theories can be
found in Ref.
\cite{Iz}. Typical example of such a theory is the model of
``nearly antiferromagnetic'' Fermi - liquid (spin - fermion model), intensively
developed by Pines and his collaborators
 \cite{Mont1,Mont2}. This model assumes
that electrons interact with spin fluctuations and the form of this interaction
(\ref{V}) is restored, as mentioned above, from the fit to NMR experiments.
\cite{MilMon}.
 This approach continues to develop rapidly, we can only mention  
the very recent papers \cite{AC,ACS,ACF}. There exist rather convincing
experimental facts supporting this mechanism of pairing \cite{Iz,ACSc}. 
At the same time interaction of the form of (\ref{V}) can be also responsible,
as we shown above, for the formation of ``dielectric'' pseudogap already at
high enough temperatures. Unfortunately, up to now there are no papers analyzing
both the pseudogap formation and superconductivity within the single approach of
spin - fermion model.

On the other hand rephrasing the quotation from Landau on Coulomb interaction,
cited in \cite{Maks}, we can only say that electron - phonon interaction was
also ``not canceled by anybody'', as a matter of fact as a possible mechanism 
of pairing in cuprates also. The review of appropriate calculations with
rather impressive results can be found in \cite{Maks,MKul} and in earlier
Ref.\cite{VLMaks}. Probably the main difficulty for the models based on
electron - phonon interaction in cuprates remains the established $d$-wave
symmetry of Cooper pairs. However, there exist interesting attempts to explain
the anisotropic pairing within the electron - phonon mechanism \cite{Maks,MKul},
as well as serious doubts in the effectiveness of spin - fluctuation
mechanism \cite{Maks}.

The author of the present review is not inclined to join either of these
positions, being convinced that these (or similar \cite{FrK})
more or less ``traditional'' mechanisms are apparently decisive for the
microscopics of HTSC. Anyhow, in the following we shall not even discuss some
other more ``radical'' approaches, e.g. those developed in different variants
and for many years models, based on the concept of Luttinger liquid \cite{PWA}, 
mainly because of the absence there of any ``stable'' results and conclusions.

Significant renormalization of electronic spectrum due to the pseudogap
formation inevitably leads to a strong change of properties of the system in
superconducting state. In the scenario of pseudogap formation due to
fluctuations of ``dielectric'' short - range order (AFM, SDW or CDW) we can
speak precisely about the influence of the pseudogap upon superconductivity.
In this formulation the analysis can be done even without assuming 
specific pairing mechanism, as it is done e.g. in the problem of the 
influence of structural disordering, impurities etc. From the physical point of
view the least justified is the used above assumption of static nature of
short - range order fluctuations, as dynamics of these fluctuations
(e.g. in the model based on (\ref{V})) is apparently decisive at low enough
temperatures (in superconducting phase), being responsible also for the
pairing mechanism itself. We deliberately use these simplification as the full
self - consistent solution of the problem within the model with dynamics seems
to be out of reach, at least at present. Besides, in the following we shall
only use the most simplified ``hot patches'' model
\footnote{This simplification is not very important. The same analysis can be
performed also in more realistic ``hot spots'' model, but this leads to much
more complicated (numerically) calculations.}. For generality we analyze both
$s$-wave and $d$-wave pairing.

So, in the spirit of the previous discussion we shall not concentrate 
on any concrete microscopic mechanism
 of pairing and
just assume the simplest possible BCS - like model, where the pairing is
due to some kind attractive ``potential'' of the following form:
\begin{equation}
V_{sc}({\bf p,p'})=V(\phi,\phi')=-Ve(\phi)e(\phi'),
\label{VV}
\end{equation}
where $\phi$ -- is polar angle defining the direction of electron momentum
${\bf p}$ in highly - conducting plane, while for  $e(\phi)$ we shall use
the simple model dependence:
\begin{equation}
e(\phi)=
\left\{
\begin{array}{ll}
1 & (\mbox{ $s$-wave pairing})\\ 
\sqrt{2}\cos(2\phi) & (\mbox{ $d$-wave pairing})
\end{array}.
\right.
\label{ephi}
\end{equation}
Attractive interaction constant $V$ is assumed, as usual, to be nonzero in some
energy interval of the width $2\omega_c$ around the Fermi level ($\omega_c$ --
is some characteristic frequency of excitations, responsible for attraction of
electrons). The model interaction of the type of (\ref{VV}) was successfully
used e.g. in Refs.\cite{Bork,Fehr} for the analysis of impurities influence on
anisotropic Cooper pairing. 

Superconducting energy gap in this case takes the following form:
\begin{equation}
\Delta({\bf p})\equiv \Delta(\phi)=\Delta e(\phi).
\label{DD}
\end{equation}
In the following, just not to make expressions too long, we shall use the
notation of $\Delta$ instead of $\Delta(\phi)$, writing the angular dependence
explicitly only when it will be necessary.

In superconducting state the perturbation theory for electron interaction with
AFM fluctuations (\ref{fluct}) should be built on ``free'' normal and anomalous
Green's functions of a superconductor:
\begin{equation}
G_{00}(\varepsilon_n{\bf p})=-\frac{i\varepsilon_n+\xi_{\bf p}}
{\varepsilon_n^2+\xi^2_{\bf p}+|\Delta|^2};\quad
F^+_{00}(\varepsilon_n{\bf p})=\frac{\Delta^*}
{\varepsilon_n^2+\xi^2_{\bf p}+|\Delta|^2}
\label{GoFo}
\end{equation}
In this case we can formulate the direct analog of approximation 
(\ref{Ansatz}) also in superconducting phase  \cite{KS2000}. 
The contribution of an arbitrary diagram of the $N$-th order in interaction
(\ref{Veff}) to either normal or anomalous Green' function consists of a
product of $N+1$ ``free'' normal $G_{0k_j}$ or anomalous $F^+_{0k_j}$ Green'
functions with renormalized (in a certain way) frequencies and gaps.
Here $k_j$ -- is the number of interaction lines surrounding the  $j$-th 
(from the beginning of the diagram) electronic line. As in normal phase the
contribution of an arbitrary diagram is defined by some set of integers
$k_j$, and each diagram with intersecting interaction lines becomes equal
to some diagram without intersections. For this last reason we can analyze
only diagrams with no intersections of interaction lines, taking all the other
diagrams into account by the same combinatorial factors $v(k)$, attributed to
interaction lines as in the normal phase. As a result we obtain an analog of
Gorkov's equations \cite{AGD}. Accordingly there appears now a system of two
recursion equations for normal and anomalous Green's functions: 
\begin{eqnarray}
G_k=G_{0k}+G_{0k}\tilde GG_k-G_{0k}\tilde F F^+_k-F_{0k}\tilde G^*F^+_k-
F_{0k}\tilde F^+G_k \nonumber\\
F^+_k=F^+_{0k}+F^+_{0k}\tilde GG_k-F^+_{0k}\tilde FF^+_k+G^*_{0k}\tilde G^*
F^+_k+G^*_{0k}\tilde F^+G_k
\label{Gork}
\end{eqnarray}
where
\begin{equation}
\tilde G=W^2v(k+1)G_{k+1};\quad \tilde F^+=W^2v(k+1)F^+_{k+1}
\label{GF}
\end{equation}
\begin{equation}
G_{0k}(\varepsilon_n{\bf p})=-\frac{i\varepsilon_n+(-1)^k\xi_{\bf p}}
{\tilde\varepsilon_n^2+\xi^2_{\bf p}+|\tilde\Delta|^2};\quad
F^+_{0k}(\varepsilon_n{\bf p})=\frac{\tilde\Delta^*}
{\tilde\varepsilon_n^2+\xi^2_{\bf p}+|\tilde\Delta|^2}
\label{GkFk}
\end{equation}
and we introduced the abovementioned renormalized frequency $\tilde\varepsilon$ 
and gap $\tilde\Delta$:
\begin{equation}
\tilde\varepsilon_n=\eta_k\varepsilon_n;\quad \tilde\Delta=\eta_k\Delta;\quad
\eta_k=1+\frac{kv_F\kappa}{\sqrt{\varepsilon_n^2+|\Delta|^2}}
\label{renpar}
\end{equation}
similar to those appearing in the analysis of a superconductor with 
impurities \cite{AGD}.

Both normal and anomalous Green's functions of a superconductor with
pseudogap are defined from (\ref{Gork}) at $k=0$ and represent the complete
sum of all diagrams of perturbation series over interaction of superconducting
electron with fluctuations of AFM short - range order.

In fact we just consider Gorkov's Green's functions averaged over an ensemble
of random (Gaussian) fluctuations of short - range order, in many respect 
analogously to the similar analysis of the case of impure superconductors
\cite{AGD}. Here we assume here the self - averaging property of 
superconducting order parameter (energy gap) $\Delta$, which allows to average
it independently of electronic Green's functions in diagrammatic series.
Usual arguments for the possibility of such independent averaging are the
following \cite{Gor,Genn,Scloc}: 
the value of $\Delta$ changes on characteristic
length scale of the order of coherence length $\xi_0\sim v_F/\Delta$ of BCS
theory, while Green's functions oscillate on much shorter length scale of the
order of interatomic distance.
 Naturally, this last assumption becomes wrong
if a new characteristic length $\xi\to\infty$  appears in electronic 
subsystem. At the same time, until this AFM correlation length
$\xi\ll\xi_0$ (i.e. when AFM fluctuations correlate on lengths smaller than
characteristic size of Cooper pairs), the assumption of self - averaging
nature of  $\Delta$ is apparently still conserved, being broken only in the
region of $\xi>\xi_0$. As a result we can use the standard approach of the 
theory of disordered superconductors (mean - field approximation in terms of
Ref.\cite{KSad}). Possible role of non - self - averaging $\Delta$ \cite{KSad} 
will be analyzed later. Note that in real 
HTSC - cuprates apparently
always we have $\xi\sim\xi_0$, 
so that they belong to the region of parameters
most difficult for theory.

\subsection {Critical Temperature and the Temperature Dependence of the Gap.}

Energy gap of a superconductor is defined by the equation:
\begin{equation}
\Delta({\bf p})=-T\sum_{\bf p'}\sum_{\varepsilon_n}V_{sc}({\bf p,p'})
F(\varepsilon_n{\bf p'})
\label{Gapeq}
\end{equation}
On flat parts of the Fermi surface the anomalous Green's function is 
determined by recursion procedure (\ref{Gork}). On the rest (``cold'') part
of the Fermi surface scattering by AFM fluctuations is absent (in our model)
and the anomalous Green's function has the form as in Eq.(\ref{GoFo}).
The results of numerical calculations of the temperature dependence of the gap
for different values of correlation length of short - range order fluctuations
$\xi$ can be found in Ref.\cite{KS2000}. These dependences are more or less of 
the usual form.

\begin{figure}
\epsfxsize=9cm
\epsfysize=8cm
\epsfbox{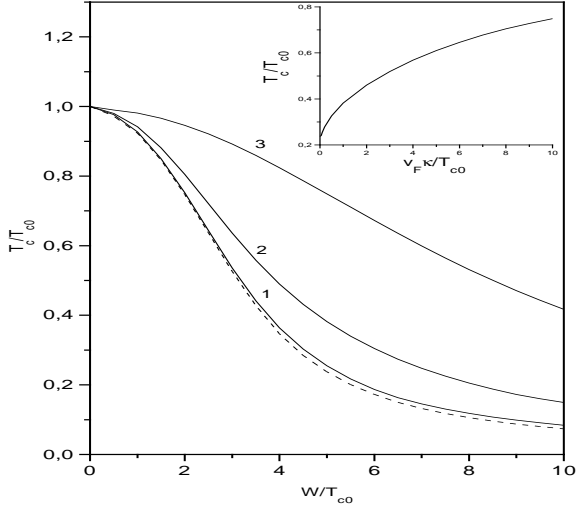}
\caption {Dependence of superconducting critical temperature on the width
of the pseudogap $W$ and correlation length of AFM fluctuations
(parameter $\kappa=\xi^{-1}$):
$\frac{v_F\kappa}{W}=0.1$ (1); \quad $\frac{v_F\kappa}{W}=1.0$ (2); 
\quad $\frac{v_F\kappa}{W}=10.0$ (3).\ 
Dashed line --- $\kappa=0$ \cite{PS}.
At the insert: dependence of $T_c$ on $\kappa$ for $\frac{W}{T_{c0}}=5$ 
\cite{KS2000}.}
\label{Tcxi}
\end{figure}

\begin{figure}
\epsfxsize=9cm
\epsfysize=8cm
\epsfbox{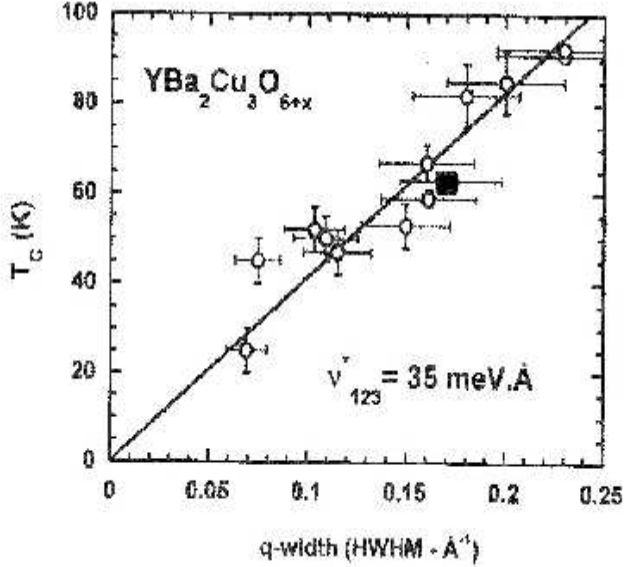}
\caption{Temperature of superconducting transition $T_c$ as a function of the
width of neutron scattering peak at ${\bf q}=(\pi/a,\pi/a)$, which is identified
with the inverse of correlation - length of short - range order fluctuations
\cite{Bour}.}
\label{xitc}
\end{figure}

Equation for the critical temperature of superconducting transition
 $T_c$ 
follows immediately from (\ref{Gapeq}) as $\Delta\to 0$. Calculated in
Ref.\cite{KS2000} dependences of $T_c$ on the width of the pseudogap
$W$ and correlation length 
(parameter $\kappa=\xi^{-1}$) are shown in 
Fig.\ref{Tcxi}, where $T_{c0}$ -- is transition temperature  in the absence
of the pseudogap. In this calculations we have, rather arbitrarily, taken
the value of $\alpha=\pi/6$,
 which is pretty close to the experimental data
of Ref.\cite{Onell}.
The general qualitative conclusion is that pseudogap suppresses 
superconductivity due to partial ``dielectrization'' of electronic spectrum
on ``hot patches'' of the Fermi surface. This suppression is maximal for
$\kappa=0$ (infinite correlation length of AFM fluctuations) \cite{PS,KSad} 
and diminishes with diminishing values of correlation length, which
qualitatively corresponds to the observed phase diagram of HTSC - systems.
As noted above, parameters of our model are to be considered as 
phenomenological. E.g. the effective width of the pseudogap $2W$ may be,
apparently, identified as $E_g$, the experimental values of which and 
dependence on doping are shown in Fig.\ref{Eg} for $YBCO$. Experimental data
on the values of correlation length $\xi$ and its temperature dependence are
rather incomplete. Indirectly this information may be extracted from NMR
experiments \cite{MilMon,Sch}. Direct neutron scattering data are not 
conclusive enough. As an example we shall mention the paper
\cite{Bour}, where the authors present some compilation of data on effective 
width of neutron scattering peak at the vector $(\pi/a,\pi/a)$ in $YBCO$ with
different oxygen content. The inverse width of this peak can be naturally
identified with the value of $\xi$. The authors of Ref.\cite{Bour} observed
an interesting correlation of superconducting critical temperature and thus
defined values of $\xi^{-1}$, which is shown in Fig.\ref{xitc}. Obviously,
this dependence is in direct qualitative correspondence with dependence of
$T_c$ on $\kappa=\xi^{-1}$, 
shown in Fig.\ref{Tcxi}. Quantitative fit of these
data to calculated dependences, following from our model, is apparently
possible, but we must be also aware of rather incomplete information on
doping dependence of parameter $W$.
In particular, it is not clear, whether there is any deep physical meaning of
$E_g\sim W$ going to zero at some ``critical'' doping as shown in fig.\ref{Eg}. 
Pseudogap influence on electronic properties may disappear also due to
correlation length $\xi$ becoming just smaller, which leads not to the
``closing'', but to the ``filling'' of the pseudogap in the density of states.
However, if we accept the data of Fig.\ref{Eg} on the width of the pseudogap
\cite{TL}, then our parameter $2W$ drops from the values of the order of
$700K$ at hole concentration $p=0.05$ to the values of the order of
$T_c\sim 100K$ near optimal doping $p=0.17$, going to zero at $p=0.19$. 
If we assume the validity of microscopic expression (\ref{dd}) for $W$,
following from the theory of ``nearly antiferromagnetic'' Fermi - liquid, 
this concentration dependence may be attributed to the appropriate dependence of
local spin $<{\bf S}_i^2>$ on $Cu$ - ion. The author is unaware of any direct 
data of this sort, but we may note Ref.\cite{JTal}, where the disappearance of
effective interaction with antiferromagnetic fluctuations at $p=0.19$ was 
demonstrated from some fit to experimental data on NMR relaxation.

Let us stress once again, that the above theoretical results are valid only
assuming self - averaging nature of superconducting order parameter (gap)
as averaged over AFM fluctuations (random mean - field approximation 
\cite{KSad}), which is apparently true for not very large values of correlation
length $\xi<\xi_0$, where $\xi_0$ --
 is the coherence length of a 
superconductor (the size of Cooper pairs at $T=0$). For $\xi\gg\xi_0$, as we
shall see below, important effects due the absence of self - averaging appear,
leading to characteristic ``tails'' in the temperature dependence of the
average gap in the region of $T_c<T<T_{co}$ \cite{KSad}.

\subsection {Cooper Instability.}

It is well known that the critical temperature may be determined also in
another way, that is from the equation for Cooper instability of the
normal phase:
\begin{equation}
1-V\chi(0,0)=0
\label{coopinst}
\end{equation}
where $\chi(0,0)$ is the generalized Cooper susceptibility:
\begin{equation}
\chi({\bf q}0;T)=-T\sum_{\varepsilon_n}\sum_{\bf p}G(\varepsilon_n{\bf p+q})
G(-\varepsilon_n{\bf p})e^2(\phi)\Gamma(\varepsilon_n,-\varepsilon_n,\bf q)
\label{chiq}
\end{equation}
Here we have to calculate the ``triangular'' vertex part
$\Gamma(\varepsilon_n,-\varepsilon_n,\bf q)$,
describing electron interaction with AFM fluctuations. For one - dimensional
analogue of our problem (and for real frequencies, $T=0$) appropriate
recursion procedure was formulated in Refs.\cite{C91}. For our two - 
dimensional model this procedure was used in calculations of optical 
conductivity, as discussed above \cite{C99}. 
This procedure is easily
generalized for the case of Matsubara frequencies \cite{KS2000}. 
Below, for definiteness, we assume 
$\varepsilon_n>0$. Then, similarly to (\ref{vertex}), we have:
\begin{eqnarray}
\Gamma_{k-1}(\varepsilon_n,-\varepsilon_n,{\bf q})=1+\nonumber\\
+W^2v(k)G_k\bar G_k
\Biggl\{1+\frac{2ikv_F\kappa}{2i\varepsilon_n-(-1)^kv_Fq-W^2v(k+1)
(G_{k+1}-\bar G_{k+1})}\Biggr\}\Gamma_{k}(\varepsilon_n,-\varepsilon_n,
{\bf q})  
\nonumber\\
\Gamma(\varepsilon_n,-\varepsilon_n,{\bf q})\equiv 
\Gamma_{0}(\varepsilon_n,-\varepsilon_n,{\bf q})
\label{Gamma}
\end{eqnarray}
where $G_k=G_k(\varepsilon_n{\bf p+q})$ and 
$\bar G_k=G_k(-\varepsilon_n,{\bf p})$
 are calculated according to (\ref{rec}).

To calculate $T_c$ we need the vertex at ${\bf q}=0$. Then
$\bar G_k=G^*_k$ and vertices $\Gamma_k$ become real, which leads to
considerable simplification of the procedure (\ref{Gamma}). 

The following exact relation holds, similar to Ward identity \cite{KS2000}:
\begin{equation}
G(\varepsilon_n{\bf p})G(-\varepsilon_n{\bf p})\Gamma(\varepsilon_n,
-\varepsilon_n,0)=-\frac{ImG(\varepsilon_n{\bf p})}{\varepsilon_n}
\label{Ward}
\end{equation}
Numerical calculations fully confirm this relation,
demonstrating the self - consistency of our recurrence relations for single -
particle Green's function and vertex part.
The relation (\ref{Ward}) leads to the equivalence of $T_c$ equation, derived
from Cooper instability, and $T_c$ equation, obtained from the linearized gap
equation, though the recurrence procedures used to derive both equations
(accounting for AFM fluctuations) seem to be quite different.

\subsection {Ginzburg-Landau Equations and Basic Properties of a 
Superconductor
 with the Pseudogap close to $T_c$.}

Ginzburg - Landau expansion for the difference of free - energy density of
superconducting and normal state can be written in standard form:
\begin{equation}
F_{s}-F_{n}=A|\Delta_{\bf q}|^2
+q^2 C|\Delta_{\bf q}|^2+\frac{B}{2}|\Delta_{\bf q}|^4,
\label{GL}
\end{equation}
where $\Delta_{\bf q}$ -- is the amplitude of Fourier - component of 
superconducting order parameter:
\begin{equation}
\Delta(\phi,{\bf q})=\Delta_{\bf q}e(\phi).
\label{FF}
\end{equation}
This expansion (\ref{GL}) is determined by loop - expansion diagrams for the
free - energy of an electron in the field of random fluctuations of order -
parameter with small wave - vector ${\bf q}$ \cite{PS}. 

Let us write the coefficients of Ginzburg - Landau expansion in the following
form:
\begin{equation}
A=A_{0}K_{A};\qquad   C=C_{0}K_{C};\qquad    B=B_0K_B,
\label{ACD}
\end{equation}
where $A_{0}$, $C_{0}$ and $B_0$ are standard expressions for these coefficients
for the case of isotropic $s$-wave pairing:
\begin{equation}
A_{0}=N_0(0)\frac{T-T_{c}}{T_{c}};\qquad
C_{0}=N_0(0)\frac{7\zeta(3)}{32\pi^{2}}\frac{v_F^2}{T_c^2};\qquad
B_0=N_0(0)\frac{7\zeta(3)}{8\pi^{2}T_c^2},
\label{ACDf}
\end{equation}
The all changes of these coefficients due to the appearance of the pseudogap
are contained in dimensionless coefficients $K_{A}$, $K_{C}$ and $K_B$.
In the absence of the pseudogap all these coefficients are equal to 1, except
$K_B=3/2$ in case of $d$-wave pairing.

Consider again the generalized Cooper susceptibility (\ref{chiq}).
Then the coefficients $K_A$ and $K_C$ can be written as \cite{KS2000}:
\begin{equation}
K_A=\frac{\chi(00;T)-\chi(00;T_c)}{A_0}
\label{Ka}
\end{equation}
\begin{equation}
K_C=\lim_{q\to 0}\frac{\chi({\bf q}0;T_c)-\chi(00;T_c)}{q^2C_0}
\label{Kc}
\end{equation}
Then all calculations can be made using the recurrence procedure (\ref{Gamma}). 

To find the coefficient $B$, in general case, we need more complicated
procedure. Significant simplifications appear if we limit ourselves, 
for terms of the order of $|\Delta_{\bf q}|^4$
 in free energy, by the usual 
case of $q=0$.  Then the coefficient $B$ may be determined directly from
the anomalous Green's function  $F$, i.e. from the recurrence procedure 
(\ref{Gork}) \cite{KS2000}. 

In the limit of $\xi\to\infty$ all coefficients of Ginzburg - Landau
expansion may be obtained analytically \cite{PS}, using the exact solution for
the Green's functions discussed above.

Ginzburg - Landau expansion defines two characteristic lengths of
superconducting state: coherence length and penetration depth (of magnetic
field).
 Coherence length at given temperature $\xi(T)$ determines the
length scale of inhomogeneities of the order parameter $\Delta$:
\begin{equation}
\xi^2(T)=-\frac{C}{A}.
\label{xii}
\end{equation}
In the absence of the pseudogap:
\begin{eqnarray}
\xi_{BCS}^2(T)=-\frac{C_{0}}{A_{0}}, \\
\xi_{BCS}(T)\approx 0.74\frac{\xi_{0}}{\sqrt{1-T/T_{c}}},
\label{xi}
\end{eqnarray}
where $\xi_{0}=0.18v_{F}/T_{c}$.\ In our model:
\begin{equation}
\frac{\xi^2(T)}{\xi_{BCS}^2(T)}=\frac{K_{C}}{K_{A}}.
\label{xiii}
\end{equation}
Appropriate dependences of $\xi^2(T)/\xi_{BCS}^2(T)$ on the width of the
pseudogap $W$ and correlation length of fluctuations (parameter $\kappa$)
,
for the case of $d$-wave pairing were calculated in Ref.\cite{KS2000} and found
to be smooth, with relatively insignificant changes of the ratio  (\ref{xiii}) 
itself.

For the penetration depth of a superconductor without the pseudogap we have:
\begin{equation}
\lambda_{BCS}(T)=\frac{1}{\sqrt{2}}\frac{\lambda_{0}}{\sqrt{1-T/T_{c}}},
\label{lamb}
\end{equation}
where $\lambda_{0}^2=\frac{mc^2}{4\pi ne^2}$ defines penetration depth at
$T=0$.\ In general case:
\begin{equation}
\lambda^2(T)=-\frac{c^2}{32\pi e^2}\frac{B}{AC}.
\label{lam}
\end{equation}
Then for our model we have:
\begin{equation}
\frac{\lambda(T)}{\lambda_{BCS}(T)}=
\left(\frac{K_{B}}{K_{A}K_{C}}\right)^{1/2}.
\label{lm}
\end{equation}
Dependences of this ratio on the pseudogap width and correlation length were 
also calculated for the case of $d$-wave pairing in Ref.\cite{KS2000} and 
found to be relatively small.

Close to $T_{c}$ the upper critical field $H_{c2}$ is determined via 
Ginzburg - Landau coefficients as:  
\begin{equation} 
H_{c2}=\frac{\phi_0}{2\pi\xi^2(T)}=-\frac{\phi_{0}}{2\pi}\frac{A}{C} ,
\label{Hc2} 
\end{equation} 
where $\phi_{0}=c\pi/e$ -- is the magnetic flux quantum. Then the slope of the
upper critical field near $T_{c}$ is given by:  
\begin{equation} 
\left|\frac{dH_{c2}}{dT}\right|_{T_c}=
\frac{24\pi\phi_{0}}{7\zeta(3)v_F^2}T_{c}
\frac{K_A}{K_C}. 
\label{dHc2}
\end{equation}
Graphic dependences of this slope $\left|\frac{dH_{c2}}{dT}\right|_{T_c}$, 
at temperature $T_{c0}$ on the effective width of the pseudogap $W$ and
correlation length parameter $\kappa$
 can also be found in Ref.\cite{KS2000}. 
The slope of the upper critical field for the case of large enough correlation
lengths drops fast with the growth of the pseudogap width. However, for short
enough correlation lengths some growth of this slope can also be observed at
small pseudogap widths. At fixed pseudogap width the slope of $H_{c2}$
grows significantly with the drop of correlation length of fluctuations.

\begin{figure}
\epsfxsize=9cm
\epsfysize=8cm
\epsfbox{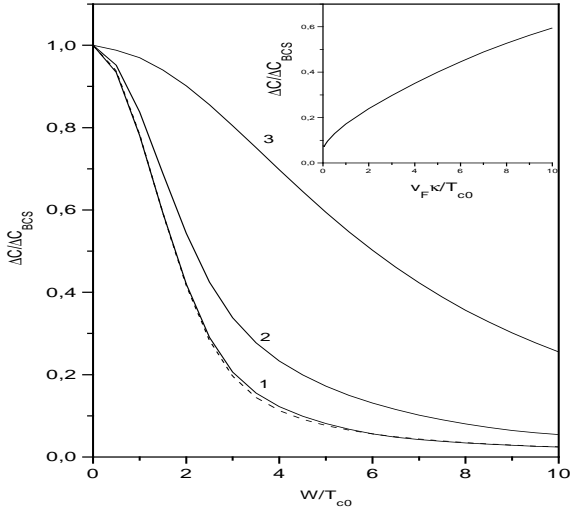}
\caption{Dependence of specific heat discontinuity on pseudogap width
$W$ and correlation length of AFM fluctuations (parameter $\kappa=\xi^{-1}$):
$\frac{v_F\kappa}{W}=0.1$\ (1);\quad $\frac{v_F\kappa}{W}=1.0$\ (2);
\quad $\frac{v_F\kappa}{W}=10.0$\ (3).
Dashed line --- $\kappa=0$ \cite{PS}.
At the insert: dependence of specific heat discontinuity on $\kappa$ for 
$\frac{W}{T_{c0}}=5$.}
\label{spht}
\end{figure}

Finally, consider the specific heat discontinuity (jump) at the transition
point:
\begin{equation} 
\frac{C_s-C_n}{\Omega}=\frac{T_c}{B}\left(\frac{A}{T-T_c}\right)^2,
\label{Cs}
\end{equation}
where $C_s,\>C_n$ -- are the specific heats in superconducting and normal
states respectively, $\Omega$ -- is the sample volume. At temperature 
$T=T_{c0}$ and in the absence of the pseudogap ($W=0$):
\begin{equation}
\left(\frac{C_s-C_n}{\Omega}\right)_{T_{c0}}=N(0)\frac{8\pi^2T_{c0}}{7\zeta(3)}.
\label{CsCn}
\end{equation}
Then the relative specific heat discontinuity in our model can be written as:
\begin{equation}
\frac{(C_s-C_n)_{T_c}}{(C_s-C_n)_{T_{c0}}}=
\frac{T_c}{T_{c0}}\frac{K_A^2}{K_B}.
\label{cscn}
\end{equation}
Appropriate dependences on effective width of the pseudogap  $W$ and correlation
length parameter $\kappa$ for the case of $d$-wave pairing are shown in 
Fig.\ref{spht}. 
It is seen that the specific heat discontinuity is rapidly
suppressed with the growth of the pseudogap width and, inversely, it grows with
for shorter correlation lengths of AFM fluctuations.

For superconductors with $s$-wave pairing all the dependences are quite
similar to that found above, the only difference is the larger scale of $W$
leading to significant changes of all characteristics of superconducting
state, corresponding to larger stability of isotropic superconductors to
partial ``dielectrization'' of electronic spectrum due to to pseudogap
formation on ``hot patches'' of the Fermi surface \cite{PS,KSad}.

All these results are in complete qualitative agreement with experimental data
on specific heat discontinuity discussed above \cite{TL,Loram,Lor}.
 As can be
seen from Fig.\ref{Cybco}, specific heat discontinuity is rapidly suppressed as
system moves to underdoped region, where the pseudogap width grows as is 
seen from the data shown in Fig.
\ref{Eg}. 

\subsection {Effects due to non-Self-Averaging Nature of the Order-Parameter.}

The above analysis of superconductivity assumed the self - averaging property
of superconducting order parameter $\Delta$. This assumption is apparently
justified in case of correlation length of fluctuations of (AFM) short - range
order $\xi$ being small in comparison with characteristic size of Cooper pairs
$\xi_0$ (coherence length of BCS theory). The opposite limit of  $\xi\gg\xi_0$ 
can be analyzed within the exactly solvable model of the pseudogap state with
$\xi\to\infty$, which was described above for the case of ``hot patches'' model 
in (\ref{fgrina}), (\ref{Rayl}) \cite{KSad}. 
 
Consider first superconductivity in a system with {\em fixed} dielectric gap
$D$ on ``hot patches'' of the Fermi surface. The problem of superconductivity
in a system with partial dielectrization of electron energy spectrum
 on
certain parts of the Fermi surface was considered previously in a number of
papers (cf. e.g. Refs.\cite{Kopaev,Ginz}), 
and in the form most useful for
our case in a paper by Bilbro and McMillan \cite{Bilb}. Below we shall use some 
of the results of this paper. Again we assume the simplest BCS - like model of
pairing interaction (\ref{VV}), (\ref{ephi}).

For the fixed value of dielectric gap  $D$ on ``hot patches'' of the Fermi
surface superconducting gap equation, defining $\Delta$ for the case of 
$s$-wave pairing, takes the following form:
\begin{equation}
1=\lambda\int_{0}^{\omega_c}d\xi \left\{\tilde\alpha\frac
{th\frac{\sqrt{\xi^2+D^2+\Delta^2(D)}}{2T}}{\sqrt{\xi^2+D^2+
\Delta^2(D)}}+(1-\tilde\alpha)\frac{th\frac{\sqrt{\xi^2+\Delta^2(D)}}{2T}}
{\sqrt{\xi^2+\Delta^2(D)}}\right\}
\label{gapsw}
\end{equation}
where $\lambda=VN_0(0)$ -- is dimensionless coupling constant of pairing
interaction, $\tilde\alpha=\frac{4\alpha}{\pi}$. 
The first term in the r.h.s.
of Eq.(\ref{gapsw}) corresponds to the contribution of ``hot'' (dielectrized or
``closed'') parts of the Fermi surface, where electronic spectrum is given by
\cite{Bilb}: $E_p=\sqrt{\xi^2_p+D^2+\Delta^2}$, while the second term gives
the contribution of ``cold'' (metallic or ``open'') parts, where the spectrum
has the usual BCS form: $E_p=\sqrt{\xi^2_p+\Delta^2}$. Eq. (\ref{gapsw})
determines superconducting gap $\Delta(D)$ at fixed value of dielectric gap
$D$, which is different from zero on ``hot patches''.

In case of $d$-wave pairing the analogous equation has the form:
\begin{equation}
1=\lambda\frac{4}{\pi}\int_{0}^{\omega_c}d\xi 
\left\{\int_{0}^{\alpha}d{\phi}e^2(\phi)\frac
{th\frac{\sqrt{\xi^2+D^2+\Delta^2(D)e^2(\phi)}}{2T}}{\sqrt{\xi^2+D^2+
\Delta^2(D)e^2(\phi)}}+\int_{\alpha}^{\pi/4}d{\phi}e^2({\phi})
\frac{th\frac{\sqrt{\xi^2+\Delta^2(D)e^2(\phi)}}{2T}}
{\sqrt{\xi^2+\Delta^2(D)e^2(\phi)}}\right\}
\label{gapdw}
\end{equation}
From these equation it can be seen that $\Delta(D)$ diminishes with the growth
of $D$, while $\Delta(0)$ coincides with the usual gap $\Delta_0$ in the absence
of dielectrization on flat parts of the Fermi surface and appearing at 
temperature $T=T_{c0}$, defined by the equation:
\begin{equation}
1=\lambda\int_{0}^{\omega_c}d\xi \frac{th\frac{\xi}{2T_{c0}}}{\xi}
\label{tc0}
\end{equation}
both for $s$-wave and $d$-wave pairing.

For $D\to\infty$ the first terms of Eqs.(\ref{gapsw}),(\ref{gapdw}) tend to 
zero, so that the appropriate equations for $\Delta_{\infty}=\Delta(D\to\infty)$ 
are:  
\begin{equation} 
1=\lambda\int_{0}^{\omega_c}d\xi
(1-\tilde\alpha)\frac{th\frac{\sqrt{\xi^2+\Delta^2_{\infty})}}{2T}}
{\sqrt{\xi^2+\Delta^2_{\infty}}}\quad \mbox{($s$-wave pairing)}
\label{gapswinf}
\end{equation}
\begin{equation}
1=\lambda\frac{4}{\pi}\int_{0}^{\omega_c}d\xi 
\int_{\alpha}^{\pi/4}d{\phi}e^2({\phi})
\frac{th\frac{\sqrt{\xi^2+\Delta^2_{\infty}e^2(\phi)}}{2T}}
{\sqrt{\xi^2+\Delta^2_{\infty}e^2(\phi)}}\quad \mbox{($d$-wave pairing)}
\label{gapdwinf}
\end{equation}
Equation (\ref{gapswinf}) coincides with gap equation for the case of $D=0$ 
with ``renormalized'' coupling constant $\tilde\lambda=\lambda(1-\tilde
\alpha)$, so that for the case of $s$-wave pairing:
\begin{equation}
\Delta_{\infty}=\Delta_0(\tilde\lambda=\lambda(1-\tilde\alpha))
\label{delsinf}
\end{equation}
and non zero superconducting gap in the limit of $D\to\infty$ appears at
$T<T_{c\infty}$
\begin{equation}
T_{c\infty}=T_{c0}(\tilde\lambda=\lambda(1-\tilde\alpha)).
\label{tcsinf}
\end{equation}
In the case of $d$-wave pairing from Eq.(\ref{gapdwinf}) we get:
\begin{equation}
T_{c\infty}=T_{c0}(\tilde\lambda=\lambda(1-\alpha_d))
\label{tcdinf}
\end{equation}
where
\begin{equation}
\alpha_d=\tilde\alpha+\frac{\sin\pi\tilde\alpha}{\pi}
\label{alphd}
\end{equation}
defines the ``effective'' size of flat parts of the Fermi surface for the
case of $d$-wave pairing. Thus, for $T<T_{c\infty}$ superconducting gap is
nonzero for any values of $D$ and drops form the value of $\Delta_0$ to 
$\Delta_{\infty}$ with the growth of $D$. For $T_{c\infty}<T<T_{c0}$  the gap
is nonzero only for $D<D_{max}$.
 Appropriate dependences of $\Delta$ on $D$ 
can easily be obtained by numerical solution of Eqs. (\ref{gapsw}) and 
(\ref{gapdw}).

In our model of the pseudogap state dielectric gap $D$ is not fixed but random
and distributed according to (\ref{Rayl}).
 The above results should be 
averaged over these fluctuations. It is important that within this model we
can directly calculate and exactly calculate superconducting gap $\Delta$,
averaged over the fluctuations of $D$: 
\begin{equation}
<\Delta> = \int_{0}^{\infty}dD{\cal P}(D)\Delta(D)=
\frac{2}{W^2}\int_{0}^{\infty}dDDe^{-\frac{D^2}{W^2}}\Delta(D)
\label{avergap}
\end{equation}
The dependences of $\Delta(D)$ described above lead to the average gap 
(\ref{avergap}) being formally nonzero up to the temperature
$T=T_{c0}$, i.e. the temperature of superconducting transition in the absence
of pseudogap anomalies. At the same time, the temperature of superconducting
transition $T_c$ in a superconductor with the pseudogap is, obviously,
smaller than $T_{c0}$ \cite{PS}.
 This paradoxical behavior of
$<\Delta>$ apparently can be explained by $D$-fluctuation induced appearance
of local regions in the sample with $\Delta\neq 0$
 (superconducting ``drops'')
in the whole temperature region of $T_c<T<T_{c0}$, with superconducting 
coherence appearing only for
 $T<T_c$. Of course, the full justification of
this qualitative picture can be obtained only after the analysis of more
realistic model with finite correlation length $\xi$ of AFM fluctuations
\footnote{Qualitatively this situation reminds the appearance of
inhomogeneous superconducting state, induced by strong fluctuations of local
electronic density of states, close to the Anderson metal - insulator
transition
 \cite{BPS,Scloc}.}. At the same time, simplicity of our ``toy''
model with  $\xi\to\infty$ allows to get an exact solution for $<\Delta>$
immediately.

To determine superconducting transition temperature $T_c$ in a sample as a 
whole we may use the standard mean - field approximation procedure to average
over random short - range order fluctuations (similar approach is used in 
the theory of impure superconductors \cite{Scloc}), which assumes the
self - averaging of superconducting gap over fluctuations of $D$
(i.e., in fact,  independence of $\Delta$ on fluctuations of $D$). 
Then the equations for this mean - field  $\Delta_{mf}$ have the following form:
\begin{equation}
1=\lambda\int_{0}^{\omega_c}d\xi \left\{\tilde\alpha\frac{2}{W^2}
\int_{0}^{\infty}dDDe^{-\frac{D^2}{W^2}}\frac
{th\frac{\sqrt{\xi^2+D^2+\Delta^2_{mf}}}{2T}}{\sqrt{\xi^2+D^2+
\Delta^2_{mf}}}+(1-\tilde\alpha)\frac{th\frac{\sqrt{\xi^2+\Delta^2_{mf}}}{2T}}
{\sqrt{\xi^2+\Delta^2_{mf}}}\right\} 
\label{gapswmf}
\end{equation}
for the case of $s$-wave pairing, and
\begin{eqnarray}
1=\lambda\frac{4}{\pi}\int_{0}^{\omega_c}d\xi\left\{\frac{2}{W^2} 
\int_{0}^{\infty}dDDe^{-\frac{D^2}{W^2}}
\int_{0}^{\alpha}d{\phi}e^2(\phi)\frac
{th\frac{\sqrt{\xi^2+D^2+\Delta^2_{mf}e^2(\phi)}}{2T}}{\sqrt{\xi^2+D^2+
\Delta^2_{mf}e^2(\phi)}}+\right.\nonumber\\
\left.+\int_{\alpha}^{\pi/4}d{\phi}e^2({\phi})
\frac{th\frac{\sqrt{\xi^2+\Delta^2_{mf}e^2(\phi)}}{2T}}
{\sqrt{\xi^2+\Delta^2_{mf}e^2(\phi)}}\right\}
\label{gapdwmf}
\end{eqnarray}
for the case of $d$-wave pairing. These equations are the limiting case
(for $\xi\to\infty$)
 of superconducting gap equations analyzed above
(\ref{Gapeq}) using the recurrence procedure (\ref{Gork}).

From (\ref{gapswmf}) and (\ref{gapdwmf}) it is easy to get the appropriate
$T_c$ - equations, giving the temperature at which the homogeneous gap
$\Delta_{mf}$ first appears everywhere in the sample. For the case of 
$s$-wave pairing we have: 
\begin{equation}
1=\lambda\left\{\tilde\alpha\frac{2}{W^2}\int_{0}^{\infty}dDDe^{-\frac{D^2}
{W^2}}\int_{0}^{\omega_c}d\xi\frac{th\frac{\sqrt{\xi^2+D^2}}{2T_c}}
{\sqrt{\xi^2+D^2}}+(1-\tilde\alpha)\int_{0}^{\omega_c}d\xi\frac{th
\frac{\xi}{2T_c}}{\xi}
 \right\}
\label{tcequs}
\end{equation}
In case of $d$-wave pairing we have just to replace $\tilde\alpha$ in 
(\ref{tcequs})
 by ``effective'' $\alpha_d$  defined in (\ref{alphd}). 
These equations for $T_c$ 
coincide with coincide with those obtained during
the microscopic derivation of Ginzburg - Landau expansion in this model
\cite{PS} and with $\xi\to\infty$ limit of equations obtained using
(\ref{Gork}), (\ref{Gapeq}).
 In general we always have $T_{c\infty}<T_c<T_{c0}$.

\begin{figure}
\epsfxsize=9cm
\epsfysize=9cm
\epsfbox{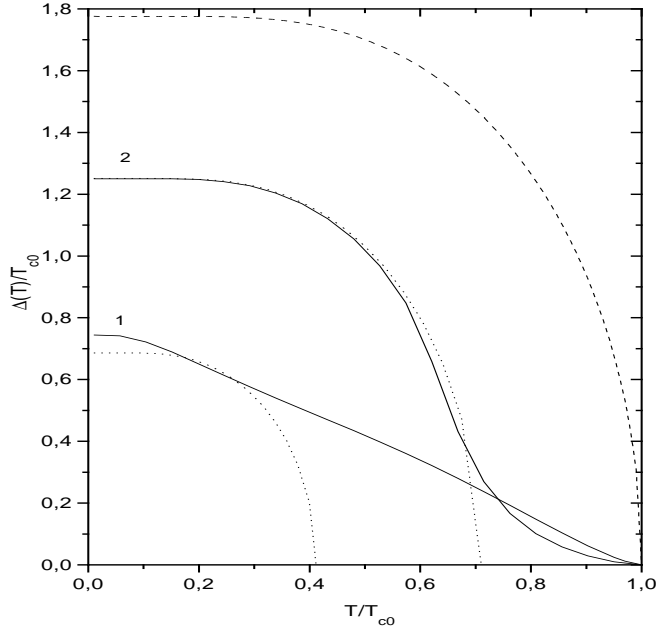}
\caption {Temperature dependences of superconducting gaps $\Delta_{mf}$ 
(dots), $<\Delta>$ (full curves) and $\Delta_{0}$ (dashed curves)
for the case of $s$-wave pairing \cite{KSad}. 
1. -- $\lambda =0.4$; $\tilde\alpha =2/3$; 
$\omega_{c}/W=3$ ($T_{c}/T_{c0}=0.42$).
2. -- $\lambda =0.4$; $\tilde\alpha =0.2$; 
$\omega_{c}/W=1$ ($T_{c}/T_{c0}=0.71$).}
\label{gaptails}
\end{figure}

Temperature dependences of average gap $<\Delta>$ and mean - field gap
$\Delta_{mf}$, obtained by numerical solution of our equations for the case of
$s$-wave pairing, are shown in Fig.\ref{gaptails}
\footnote{In case of $d$-wave pairing temperature dependences of $<\Delta>$ 
and $\Delta_{mf}$ are qualitatively similar to the case of $s$-wave pairing.}. 
We see that $\Delta_{mf}$ goes to zero at $T=T_c<T_{c0}$, while $<\Delta>$ is 
nonzero up to $T=T_{c0}$ and characteristic ``tails'' in the temperature
dependence of $<\Delta>$  in the region of $T_c<T<T_{c0}$, in our opinion,
correspond to the abovementioned picture of superconducting ``drops'' appearing
in this temperature interval, while superconductivity in a sample as a whole
is absent. Note that temperature dependences of $<\Delta(T)>$ shown in
Fig.\ref{gaptails} are quite similar to those observed in underdoped cuprates
via ARPES \cite{Nrm} or specific heat measurements \cite{Lor}, and shown in
Figs. \ref{DcT}, \ref{ARPGAP}, 
if we assume that $T_c$ observed in these 
samples corresponds to our mean - field $T_c$, while ``drops'' with
$<\Delta>\neq 0$  exist in the region of $T>T_c$ up to $T_{c0}$, which is much
higher than $T_c$.
 Non self - averaging nature of the gap in our model is
also reflected in the fact that in the whole temperature region of
$T<T_{c0}$ gap dispersion $<\Delta^2>-
<\Delta>^2$ is different from zero,
circumstantially confirming the proposed qualitative picture of ``drops''.
Of course, full justification of this picture is possible only with the
account of finite values of $\xi$.
 Note that the value of $T_{c0}$ is not very
well defined from the experimental point of view. Speaking above about 
superconducting transition temperature in the absence of the pseudogap
we assumed the value of $T_{c0}$ of the order of $T_c$ 
observed for optimal
doping. However, to compare the results shown in Fig.\ref{gaptails} with data
of Figs.\ref{DcT}, \ref{ARPGAP} we need the values of $T_{c0}$ much higher than
optimal $T_c$. in this sense the described picture of superconducting ``drops''
induced by ``dielectric'' gap fluctuations becomes not so very different from
precursor pairing picture of superconducting pseudogap scenario.
At the same time it is clear that the most part of data shown in Figs.
\ref{DcT}, \ref{ARPGAP} may be explained by the existence of the pseudogap
of dielectric nature, not by just discussed non self - averaging nature of
superconducting $\Delta$. Note, for example, that superconducting gap obtained
from tunneling data of Refs. \cite{Kras1,Kras2} 
(appearing on the ``background'' of much wider pseudogap) demonstrates
``normal''
 temperature behavior going to zero at $T=T_c$. 

Despite the fact that superconductivity in a sample as a whole for
$T_c<T<T_{c0}$ is absent, the existence in this region of nonzero average
$<\Delta>$ leads to anomalies in  
a number of observables, e.g. in tunneling
density of states or spectral density measured via ARPES \cite{KSad}.

In particular, the tunneling density of states for the case of
$d$-wave pairing has the following form \cite{KSad}:
\begin{eqnarray}
\frac{N(E)}{N_0(0)}=\frac{4}{\pi}\frac{2}{W^2}\int_{0}^{\infty}dDD
e^{-\frac{D^2}{W^2}}\left\{\int_{0}^{\alpha}d\phi\frac{|E|}
{\sqrt{E^2-D^2-\Delta^2(D)e^2(\phi)}}\theta(E^2-\Delta^2(D)e^2(\phi)-D^2)+
\right.\nonumber\\ 
\left.+\int_{\alpha}^{\pi/4}d\phi\frac{|E|}{\sqrt{E^2-
\Delta^2(D)e^2(\phi)}}\theta(E^2-\Delta^2(D)e^2(\phi))\right\}
 \nonumber\\
\label{dosdw}
\end{eqnarray}
Assuming the self - averaging nature of the gap we have
$\Delta=\Delta_{mf}$ and independent of $D$.
 In this case the width of
superconducting pseudogap in the density of states is of the order of
$\Delta_{mf}$ and appropriate contribution disappears for
$T\to T_c$, only pseudogap due to AFM fluctuations remains. In reality we have
in (\ref{dosdw}) $\Delta=\Delta(D)$, which is determined from
 Eq.(\ref{gapdw}). 

\begin{figure}
\epsfxsize=9cm
\epsfysize=9cm
\epsfbox{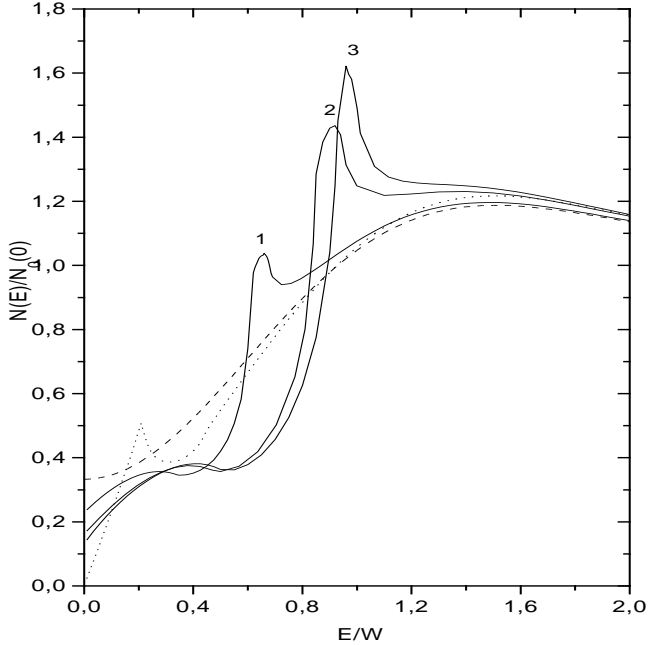}
\caption {Density of states in case of $d$-wave pairing \cite{KSad}.
$\lambda =0.4$; $\tilde\alpha =2/3$; $\omega_{c}/W=5$ 
($T_{c}/T_{c0}=0.48$, $T_{c\infty}/T_{c0}\sim 10^{-18}$),
$T/T_{c0}=$: 1. -- 0.8; 2. -- 0.48; 3. -- 0.1.
Dotted line shows the mean - field density of states 
$N_{mf}(E)$ for $T/T_{c0}=0.1$.
Dashed line -- density of states with pseudogap for $T>T_{c0}$.}
\label{dosdex}
\end{figure}

Tunneling density of states for the case of $d$-wave pairing is shown in 
Fig.\ref{dosdex}.  
We see the significant difference of the exact density of states from that
obtained in the mean - field approximation, which is due to fluctuations of
superconducting gap (superconducting ``drops'') induced by fluctuations of
AFM short range order. The exact density of states in fact does not ``feel''
superconducting transition in a sample as a whole at $T=T_c$. 
Characteristic scale of superconducting gap (pseudogap) in tunneling density
of states becomes of the order of $\Delta_0$, not $\Delta_{mf}$,
 as in the
mean - field approximation over fluctuations of short - range order.
Superconducting contributions to the density of states become observable
already for $T=T_{c0}>T_c$.
 
These results may, in principle, explain unusually high values of the ratio
$2\Delta/T_c$,
 observed in some tunneling experiments in underdoped cuprates
\cite{Renn,biggaps}, as well as in ARPES \cite{biggp}. 
The presence of
superconducting ``drops'' may also explain unusual diamagnetism of these
systems sometimes observed above $T_c$ \cite{varlm}.

\section {Conclusion. Problems for the Future.}

Finally let us discuss some general conclusions and remaining problems.
Scenario of pseudogap formation, based on the picture of fluctuations of
AFM (SDW,CDW) short - range order leads to general qualitative agreement with
main experimental facts on pseudogap in HTSC - cuprates.
Our theoretical analysis was intentionally semi - phenomenological and we
just ``modeled'' pseudogap anomalies of electronic spectrum, described by
parameters of effective pseudogap width $W$ and correlation length
$\xi$, which, in principle, may be determined from the experiment.
Both ``hot spots'' and ``hot patches'' models can be solved ``nearly exactly''
\cite{Sch,KS}, which allows significant progress in the analysis of pseudogap
influence upon superconductivity. Let us stress once again that it is not very
important whether we use AFM - fluctuations model (as most popular in the
literature) of the model of short -rang order fluctuations of some other
nature, leading to partial dielectrization of the spectrum (such as CDW or 
structural).

Deficiencies of our models are due to simplifying assumptions necessary to obtain
these ``nearly exact'' solutions. The main of these are the static
approximation and assumption of Gaussian nature of short - range order
fluctuations. The account of dynamic nature of fluctuations is absolutely
necessary for low enough temperatures, particularly in superconducting phase
where pairing itself may be due dynamics of these fluctuations
\cite{Mont1,Mont2}. We feel, however, that simplified analysis given above
may in fact describe the most important effects of pseudogap renormalization
of electronic spectrum (pseudogap formation on ``hot'' parts of the Fermi
surface) and its influence upon superconductivity. The account of dynamics of
spin fluctuations will inevitably lead us outside the simple phenomenology of
BCS model. Assumption of Gaussian statistics of fluctuations also may be
justified only for high enough temperatures and not very close to the line of
antiferromagnetic instability. However, even this simplest approximation allows
qualitative description of almost all basic anomalies due to pseudogap
formation. Without this assumption the simple and physically appealing
structure of our equations is destroyed, but only this structure allows us
to analyze superconducting phase in a relatively simple manner.

The extension of our semi - phenomenological approach can be apparently done
within the framework of full microscopic analysis of the problem, e.g. on the
basis of the Hubbard model. Such attempts were made in a number of papers, 
e.g. in Refs. \cite{Kam,Benn,Dei} cited above, as well as in Refs.
\cite{SLGB,VilTre}. Though many qualitative conclusions of these papers
coincide or are close to those discussed above, the analysis is in most cases 
restricted to first diagrams of perturbation series \cite{Vil,Month} and some
kind of self - consistent procedure on this basis. In this case there is no
big problem also to take into account also the dynamics of spin fluctuations.
However, any generalization to higher order contributions is difficult enough.

Microscopic justification of the existence of the wide region of the phase
diagram with well developed fluctuations of AFM (SDW) short - range order is,
of course, of principal importance. It should be noted that this region of
``critical'' fluctuations may be wide enough just due to the low dimensionality
(quasi two - dimensional nature) of HTSC - systems. For example, the region
of superconducting critical fluctuations in HTSC - cuprates may reach the
width of some tens of degrees $K$ \cite{Jun}. As AFM fluctuations are 
characterized by an energy scale which is an order of magnitude higher, the
existence of the ``critical'' region with the width of some hundreds degrees
does not seem improbable. However, the microscopic justification of such
picture is still absent.

Probably the main qualitative conclusion from our theoretical picture
concerns the old discussion about Fermi - liquid or non Fermi - liquid nature
of electronic spectrum in HTSC - cuprates. According to our simple models the
electronic spectrum (spectral density) is Fermi - liquid like only on ``cold''
parts of the Fermi surface (in the vicinity of Brillouin zone diagonals),
while on ``hot'' parts the spectral density is in general non Fermi - liquid 
like due to strong scattering on AFM fluctuations \cite{Sch,KS}. 
The presence or absence of Fermi - liquid behavior essentially depends on the
value of correlation length of short - range order $\xi$ 
(Cf. Refs.\cite{C91a,McK}). 
In a recent paper \cite{VarAbr} this picture was
criticized on the basis of successful
 enough fit of experimental ARPES data
of Refs.\cite{Kami,Vall,Valla} for optimally doped $Bi-2212$ to dependences
of the ``marginal'' Fermi - liquid theory. Particularly, in Ref.\cite{VarAbr} it
was claimed that Fermi - liquid like behavior of the quasiparticle damping is
absent everywhere on the Fermi surface and there is no significant momentum
dependence of this damping. In fact the authors of Ref. \cite{VarAbr} had to
introduce the significant momentum dependence of static scattering rate
attributed to anisotropic impurity scattering, which was necessary to fit data, 
even for optimally doped system. Linear dependence of damping rate on the
energy of quasiparticles, which is observed in HTSC - cuprates (and postulated
in ``marginal'' Fermi - liquid theory), can not disprove the usual Fermi -
liquid theory, because the standard quadratic dependence of da[21~mping may be
observable only in some very narrow energy interval close to Fermi level,
which is apparently smaller than typical precision of ARPES experiments.
Rather detailed discussion of this point of view can be found in
recent reviews \cite{Maks,VLMaks}.
 It must be stressed that the existence
of anomalous scattering by AFM fluctuations (with wave vectors of the order of
the vector of antiferromagnetism) is beyond any doubts
\footnote { Note an interesting recent paper \cite{LiWu}, where rather good
description of unusual transport properties of HTSC - oxides was obtained
within the model of simplest possible 
Boltzmann like scattering by 
spin fluctuations, assuming that the contribution of ``hot'' parts of the
Fermi surface to electron transport is just excluded.}.

Our theoretical discussion also used rather traditional scheme of averaging
over the random field of AFM (SDW,CDW) fluctuations, which assumes the average
spatial homogeneity of the system. At the same time a number of experimental
data and theoretical models stress the possible phase separation in some
HTSC - cuprates (especially in underdoped region) \cite{Nag,Zaan}. This phase
separation takes place on microscopic scales and the system consists of
``metallic'' (superconducting) and ``dielectric'' (antiferromagnetic) 
``stripes'' with the width of the order of few interatomic distances.
Naturally this situation can hardly be described using standard methods used
above, while in the opinion of many authors these ``stripes'' may be of primary
importance for the physics of HTSC. We shall only note, that the qualitative
picture of the random field of AFM (SDW,CDW) fluctuations introduced above also
implicitly assumes the appearance of effectively antiferromagnetic regions of
the size of the order of $\sim\xi$ intermixed with the regions of the same
size, where AFM order is destroyed. In this sense the picture used above seems
to be not drastically different from the concept of phase separation 
(``stripes''), being different mainly by the smoothness of borders between
the ``metallic'' and ``non - metallic'' regions. Note also that the formalism 
of the number of models of phase separation is based upon the idea of the
closeness of the system to some kind of CDW instability \cite{DiCas,Cast,Capr} 
and explicitly introduces scattering by fluctuations with special wave vectors. 

It is well known that a lot of anomalies is observed in HTSC - systems
under strong structural disordering \cite{Scloc}. The question of the
importance of structural disordering in the pseudogap state is almost not
investigated. We may mention only few papers, where the effects of random
impurities were studied \cite{Tal1,Tal2}. From these preliminary results it may
be concluded that controlled disordering may be rather informative method of
studies of the pseudogap state. Theoretically these problems are not studied 
at all.

The studies of pseudogap anomalies in HTSC - systems can lead also to rather
important conclusions with respect to possible applications.
We have seen that pseudogap formation in some sense suppresses 
superconductivity and, in fact, such important characteristics of 
superconductors as critical current and critical magnetic fields are maximal
not at optimal doping  $p_0\approx 0.15-0.17$, corresponding to maximal
$T_c$, but in the vicinity of the ``critical'' doping $p_c\approx 0.19$ 
\cite{TWL}. This fact may be of some significance for optimization of
technical superconductors made of HTSC - cuprates.

The author expresses his  gratitude to Dr. E.Z.Kuchinskii for his collaboration 
in some of the basic papers used in theoretical part of this review.

This work was supported in part by the grants of the Russian Foundation of 
Basic Research 99-02-16285 and CRDF No. REC-005,
 as well as by the projects
108-11(00)-$\Pi$ of the State Program on ``Statistical Physics'' and 
96-051 of the State Program on HTSC.

\end{document}